\documentclass[sn-chicago]{sn-jnl}

\usepackage{lineno}

\jyear{2023}%
\usepackage{booktabs}

\usepackage{rotating} 

\begin{document}
\title[Simulation of geologic CO$_2$ storage: Value of Local Data and Predictability]{Direct Comparison of Numerical Simulations and Experiments of CO$_2$ Injection and Migration in Geologic Media: Value of Local Data and Predictability}

\author[1,2]{\fnm{Llu{\'i}s} \sur{Sal{\'o}-Salgado}}\email{lsalo@mit.edu}

\author[3]{\fnm{Malin} \sur{Haugen}}\email{Malin.Haugen@uib.no}

\author[3]{\fnm{Kristoffer} \sur{Eikehaug}}\email{Kristoffer.Eikehaug@uib.no}

\author*[3,4]{\fnm{Martin} \sur{Fernø}}\email{Martin.Ferno@uib.no}

\author*[5,4]{\fnm{Jan M.} \sur{Nordbotten}}\email{Jan.Nordbotten@uib.no}

\author*[1,2]{\fnm{Ruben} \sur{Juanes}}\email{juanes@mit.edu}

\affil*[1]{\orgdiv{Department of Civil and Environmental Engineering}, \orgname{Massachusetts Institute of Technology}, \orgaddress{\city{Cambridge}, \postcode{02139}, \state{MA}, \country{USA}}}

\affil[2]{\orgdiv{Earth Resources Laboratory, Department of Earth, Atmospheric and Planetary Sciences}, \orgname{Massachusetts Institute of Technology}, \orgaddress{\city{Cambridge}, \postcode{02139}, \state{MA}, \country{USA}}}

\affil[3]{\orgdiv{Department of Physics and Technology}, \orgname{University of Bergen}, \orgaddress{\city{Bergen}, \postcode{5020}, \country{Norway}}}

\affil[4]{\orgname{Norwegian Research Center}, \orgaddress{\street{Postboks 22 Nygårdstangen} \city{Bergen}, \postcode{5020}, \country{Norway}}}

\affil[5]{\orgdiv{Center for Modeling of Coupled Subsurface Dynamics, Department of Mathematics}, \orgname{University of Bergen}, \orgaddress{\city{Bergen}, \postcode{5020}, \country{Norway}}}


\abstract{\textbf{Purpose:} The accuracy and robustness of numerical models of geologic CO$_2$  sequestration are almost never quantified with respect to direct observations that provide a ground truth. Here, we conduct CO$_2$ injection experiments in meter-scale, quasi-2D tanks with porous media representing stratigraphic sections of the subsurface, and combine them with numerical simulations of those experiments.

\textbf{Goals:} We evaluate (1) the value of prior knowledge of the system, expressed in terms of ex-situ measurements of the tank sands’ multiphase flow properties (local data), to obtain an accurate simulation; and (2) the predictive capability of the matched numerical models, when applied to different settings. 

\textbf{Methods:} We match three different simulation models —each with access to an increasing level of local data—to a CO$_2$ injection experiment in tank 1 (89.7$\times$47$\times$1.05 cm). Matching is based on a quantitative comparison of CO$_2$ migration at different times from timelapse image analysis. Next, we simulate a different injection scenario in tank 1, and, finally, a different injection scenario in tank 2 (2.86$\times$1.3$\times$0.019 m), which represents an altogether different stratigraphic section.

\textbf{Results and conclusion:} Our models can qualitatively match the CO$_2$ plume migration and convective mixing of the experimental truth. Quantitatively, simulations are accurate during the injection phase but their performance decreases with time. Using local data reduces the time required to history match. The predictive capability of matched models, however, is found to be similar. The sand-water-CO$_{2(\text{g})}$ system is very sensitive to effective permeability and capillary pressure changes; where heterogeneous structures are present, accurate deterministic estimates of CO$_2$ migration are unlikely.

%
}

\keywords{CO$_2$ storage, geologic carbon sequestration, two-phase flow, numerical simulations, history matching, FluidFlower}



\maketitle

\section{Introduction}\label{sec1}
CO$_2$ capture and subsequent geologic carbon sequestration (GCS) is a climate-change mitigation technology that can be deployed at scale to oﬀset anthropogenic CO$_2$ emissions during the energy transition~\citep{marcucci2017,easac2018,celia2021,ipcc2022}. In GCS, reservoir simulation is the primary tool to assess subsurface CO$_2$ migration, which is necessary to understand and manage geologic hazards such as fault leakage~\citep[e.g.,][]{caine1996,ingram1999,nordbotten2012book,zoback2012,juanes2012,jung2014,vilarrasa2015,salo2023} and induced seismicity~\citep[e.g.,][]{cappa2011,zoback2012,juanes2012,ellsworth2013,verdon2013,alghannam2020,hager2021}. In response to the inherent uncertainties associated with modeling and simulation of CO$_2$ storage \citep{nordbotten2012}, building confidence in simulation models requires calibration (or, synonymously, history matching), a process that involves updating the reservoir model to match field observations as they become available~\citep{oliver2011}.

History matching is an ill-posed inverse problem~\citep{oliver2011}. This means that multiple solutions (i.e., parameter combinations) exist that approximate the data equally well. Automated techniques such as Markov chain Monte Carlo, randomized maximum likelihood or ensemble-based methods can be used to quantify uncertainty in history-matched models, especially in combination with surrogate models to reduce forward model computational time~\citep[see][forthcoming, and references therein]{aanonsen2009,oliver2011,jagalur2018,jin2019,liu2020,santoso2021,landa2023}. In practice, however, it may be difficult to ensure that the chosen simulation model provides the best possible forecast. This is due to different subsurface conditions, the inability to include all sources of uncertainty in the models, incomplete field data and time limitations.

In the laboratory, intermediate-scale ($\sim$meter) experiments have been used to study the physics of petroleum displacement~\citep[e.g.,][]{gaucher1960,brock1991,cinar2006} and contaminant transport~\citep[e.g.,][]{silliman1987,wood1994,lenhard1995,fernandez2004}. Similar 2D and 3D flow rigs have recently been applied to CO$_2$ storage, providing a link between core-scale measurements and field observations:~\cite{neufeld2010} studied the scaling of convective dissolution and found it to be an important mechanism in the long-term trapping of injected CO$_2$ in an idealized site.~\cite{wang2010} used a 3D setup to investigate the ability of electrical resistivity tomography to identify localized leaks.~\cite{trevisan2014,trevisan2017} focused on the impact of structural and residual trapping. In homogeneous sands, they found that previous trapping models, such as the~\cite{land1968} model, can approximate the residually trapped gas saturation (R$^2 > 0.6$). Studying an heterogeneous aquifer characterized by a log-normal distribution of six different sand facies, they report that trapping efficiency increased significantly due to structural trapping.~\cite{askar2021} used a $\sim$8 m-long tank to test a framework for GCS monitoring of CO$_2$ leakage. These studies employed homogeneous glass beads or sands, or focused on heterogeneities in the aquifer layer; structural complexity was minimal.

In this paper, we use quasi-2D, intermediate-scale experiments of CO$_2$ storage to evaluate, quantitatively, the predictive capability of history-matched simulation models against well-defined spatial data. An attempt was made to recreate realistic basin geometries, including stacking of storage reservoirs, faults, caprock and overburden. We simulate each of the three presented experiments with three different simulation models. This allows us to assess (1) the value of local information of the system, expressed in terms of sand petrophysical measurements, during history matching, and (2) transferability or predictive capability of our matched simulation models, when tested against a different experiment.

\section{Physical Experiments} \label{sec:experiments}
\begin{figure}[h]%
\centering
\includegraphics[width=1\textwidth]{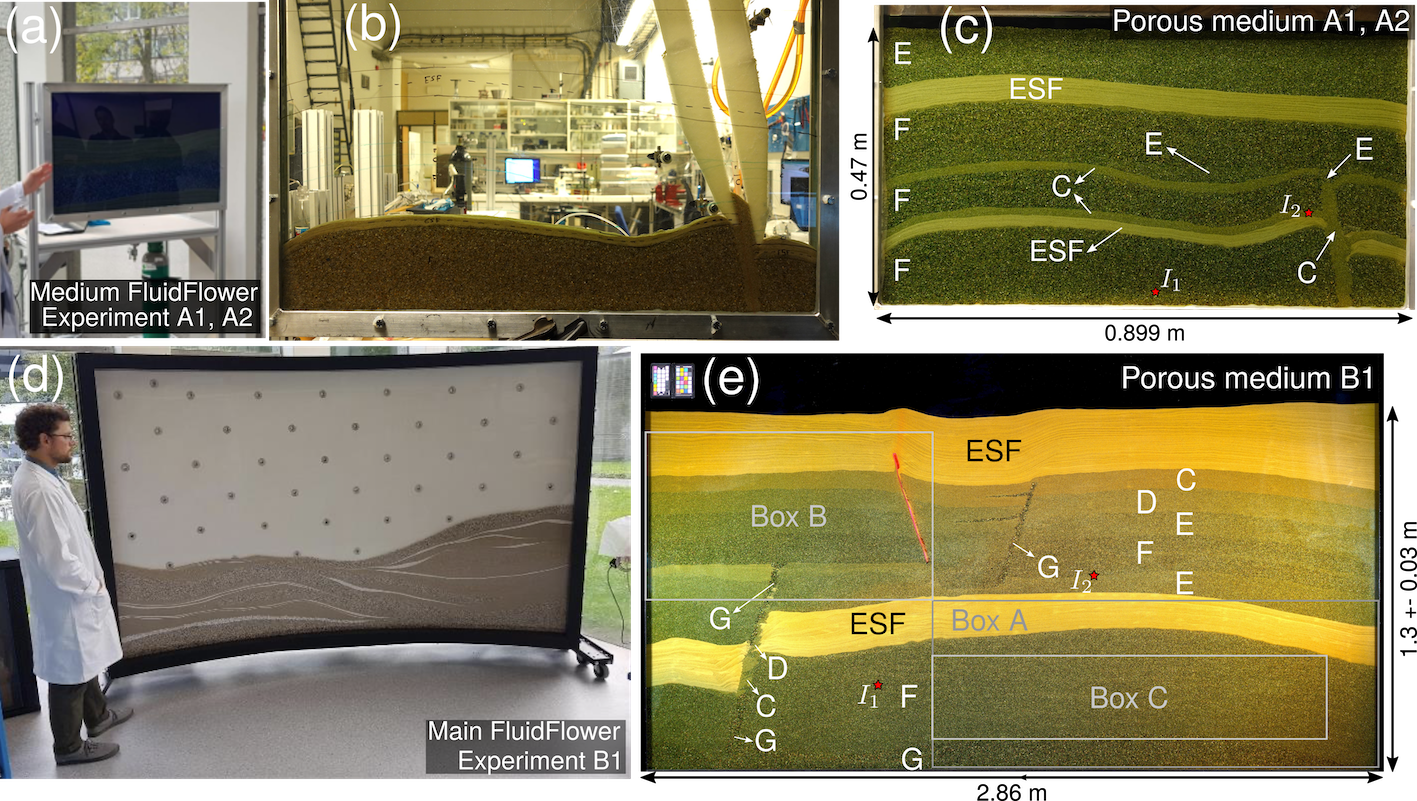}
\caption{Overview of the FluidFlower rigs and porous media used in the physical experiments. \textbf{a} Medium FluidFlower rig (tank 1). \textbf{b} Snapshot during sand pouring to build the porous medium used in experiments A1 and A2 in tank 1~\citep{haugen2023}. \textbf{c} Front view of porous medium in tank 1, with lithologies in white and injector location shown with a red star. The length and height correspond to the porous medium. Note the fixed water table at the top. \textbf{d} Overview of the main FluidFlower rig (tank 2), showing the back panel with sensor network. \textbf{e} Porous medium in tank 2, used for experiment B1, with lithologies in white. Location of injectors and boxes A, B and C for analysis are shown with a red star and gray boxes, respectively. Length and variable height correspond to the porous medium.}
\label{fig:experimentalSetup}
\end{figure}

The physical experiments of CO$_2$ injection are conducted using the {\em FluidFlower} rigs. These rigs are meter-scale, quasi-2D tanks with transparent Plexiglass panels designed and built in-house at the University of Bergen (Fig.~\ref{fig:experimentalSetup}). Here, we used two tanks, with dimensions $89.9\times47\times1.05$ cm and $2.86\times1.3\times0.019$ m (referred herein to as tank 1 and tank 2, respectively) . Different geologic settings are constructed by pouring unconsolidated sands with desired grain sizes in the water-saturated rigs. The rigs have multiple ports which allow flushing out fluids after a given CO$_2$ injection, such that multiple injections can be conducted in the same setting. The location of the ports can be adjusted to accommodate different injection scenarios. A variety of techniques have been developed by UiB engineers in order to build complex structures such as folds and faults. 

Below, we summarize the petrophysical measurements, experimental setup, geologic model/porous media construction and experimental schedule. Details on the conceptualization of the FluidFlower rigs and technical information are given in~\cite{ferno2023,eikehaug2023}, this issue, while the full description of the physical experiment in tank 1 and ex-situ measurements are provided by~\cite{haugen2023}, this issue. Further details on the experiment in tank 2 is provided by~\cite{flemisch2023}, this issue.

\subsection{Sand petrophysical properties} \label{sec:measurements}
Measurements on the employed Danish quartz sands were conducted using specialized equipment to determine average grain size ($d$), porosity ($\phi$), permeability ($k$), capillary entry pressure ($p_\text{e}$) and drainage and imbibition saturation endpoints (denoted as connate water saturation, $S_\text{wc}$, and trapped gas saturation, $S_\text{gt}$). The methodology is described by~\cite{haugen2023}, this issue, and obtained values are provided in Tab.~\ref{tab:measurements}.

\begin{table}[h] 
\centering
\caption{Petrophysical properties for used quartz sands, as obtained from local, ex-situ measurements. Measured gas column heights for sands E-G were 0, so $p_\text{e}$ could not be directly quantified. Methodology is provided by~\cite{haugen2023}.}
\begin{tabular}{ccccccc}
\toprule
Sand type & $d$ [mm] & $\phi$ [-] & $k$ [D] & $p_\text{e}$ [mbar] & $S_\text{wc}$ & $S_\text{gt}$ \\ \midrule
  ESF & 0.2 & 0.435 & 44 & 15 & 0.32 & 0.14\\
  C & 0.66 & 0.435 & 473 & 3 & 0.14 & 0.1 \\
  D & 1.05 & 0.44 & 1110 & 1 & 0.12 & 0.08 \\
  E & 1.45 & 0.45 & 2005 & - & 0.12 & 0.06 \\
  F & 1.77 & 0.44 & 4259 & - & 0.12 & 0.13 \\
  G & 2.51 & 0.45 & 9580 & - & 0.1 & 0.06 \\ \bottomrule
\end{tabular}
\label{tab:measurements}
\end{table} 

\subsection{Experimental setup} \label{sec:experimentSetup}
The front and back panels of the FluidFlower are mounted on a portable aluminum frame, such that boundaries are closed on the sides and bottom (no flow). The top surface is open and in contact with fluctuating atmospheric pressure (Fig.~\ref{fig:experimentalSetup}). A fixed water table above the top of the porous medium was kept throughout the experiments conducted here. The experimental setup incorporates mass flow controllers to inject gaseous CO$_2$ at the desired rate, and a high-resolution photo camera with time-lapse function~\citep[this issue]{haugen2023}. 

Experiments were conducted in 2021 and 2022 in Bergen (Norway) at room temperature ($\approx 23$ $^\circ$C) and fluctuating atmospheric pressure. Temperature changes were minimized as much as possible, but maintaining a constant temperature was not possible due to available laboratory space. The fluids and sands were set in the FluidFlowers using the following procedure:
\begin{enumerate}
    \item The silica sands are cleaned using an acid solution of water and HCl to remove carbonate impurities.
    \item The FluidFlower rig is filled with deionized water.
    \item Sands are manually poured into the rig using the open top boundary, in order to construct the desired porous medium.
    \item A pH-sensitive, deionized-water solution containing bromothymol blue, methyl red, hydroxide and sodium ions is injected through multiple ports until the rig is fully saturated. This enables direct visualization of CO$_2$ gas (white), dissolved CO$_2$ (orange to red), and pure water (blue).
    \item 5.0 purity (99.999\%) CO$_2$ is injected as gaseous phase at the desired rate.
    \item After the injection phase, injection ports are closed and CO$_2$ migration continues.
    \item Once the experiment is finished, the rig can be flushed with deionized water and the process can start again from step 4.
\end{enumerate}
Full details on the fluids are given in~\cite{ferno2023,eikehaug2023}, this issue. Below, we refer to the pH-sensitive solution in the rigs as ``dyed water".

\subsection{Porous media geometries}
The geometries of the porous media used in this paper aim to recreate the trap systems observed in faulted, siliciclastic, petroleoum-bearing basins around the world, given the geometrical constraints of the FluidFlowers and manual sand pouring~\citep[this issue]{ferno2023,eikehaug2023}. Features such as folds, faults and unconformities were built in both tank 1 and 2. The construction of faults, shown in Fig.~\ref{fig:experimentalSetup}b and detailed in~\cite{haugen2023}, requires a minimum thickness; hence, our fault structures are thicker than natural faults with the same displacement~\citep{chi09}. Fine sands ($d\approx 0.2$ mm) are used as sealing or caprock formation.

The geometry in tank 1 (Fig.~\ref{fig:experimentalSetup}c) contains three main high-permeability reservoirs (F sand). The bottom and middle F sand are separated by a seal (ESF sand), while the middle and top are separated by the C sand and connected through a higher permeability fault (refer to sect.~\ref{sec:measurements} for pertrophysical properties). The fault separates the bottom section in two compartments. The bottom and top F sand provide anticlinal traps for the CO$_2$ to accumulate in.

The geometry in tank 2 (Fig.~\ref{fig:experimentalSetup}e) was specifically motivated by the structure of North Sea reservoirs and petroleum basins. From bottom to top, it contains two sections of decreasing-permeability reservoirs capped by two main sealing layers. A fault separates the bottom section in two compartments, while two faults separate the top section in three compartments. Each fault has different petrophysical properties: The bottom fault is a heterogeneous structure containing ESF, C, D, F and G sands, the top-left fault is an impermeable structure made of silicone and the top-right fault is a conduit structure containing G sand.

\subsection{Experimental injection schedule} \label{sec:experimentSchedule}
The injection schedule for experiments in tank 1 is provided in Tab.~\ref{tab:experimentSchedule}. Injection ports have an inner diameter of 1.8 mm.

\begin{table}[h] 
\centering
\caption{Schedules for the three CO$_2$ injection experiments simulated in this work. $I_\text{R}$ is injection rate, while $I_i$ denotes injector (port) number. A five-minute ramp-up and ramp-down was applied in experiments A1 and A2 in tank 1. Total duration of conducted experiments and simulations is 48h (A1), 5h (A2) and 120h (B1). Location of injection wells is provided in Fig.~\ref{fig:experimentalSetup}.}
\begin{tabular}{p{0.2\textwidth} p{0.15\textwidth} | p{0.1\textwidth} p{0.1\textwidth} | p{0.1\textwidth} p{0.1\textwidth}} \toprule
Experiment A1 & & A2 &  & B1 & \\ \midrule
$I_\text{R}$ [ml/min] & $t$ [hh:mm:ss] & $I_\text{R}$ & $t$ & $I_\text{R}$ & $t$ \\ \midrule
 0.1 ($I_1$) & 00:00:00 & 0.1 ($I_1$) & 00:00:00 & 10.0 ($I_1$) & 00:00:00\\
 2.0 & 00:05:00 & 2.0 & 00:05:00 & 10.0 & 05:00:00 \\
 2.0 & 00:50:00 & 2.0 & 04:43:44 & 0.0 & 05:00:01 \\
 0.0 & 00:55:00 & 0.0 & 04:48:33 & 10.0 ($I_2$) & 02:15:00 \\
 0.1 ($I_2$) & 01:09:11 & 0.0 & 05:00:00 & 10.0 & 05:00:00 \\
 2.0 & 01:14:11 & & & 0.0 & 05:00:01 \\
 2.0 & 02:29:11 & & & 0.0 & 120:00:00 \\
 0.0 & 02:34:00 & & & & \\
 0.0 & 48:00:00 & & & & \\ \bottomrule
\end{tabular}
\label{tab:experimentSchedule}
\end{table} 

\section{Numerical simulations} \label{sec:numerical}
\subsection{Model setup} \label{sec:setup}
The isothermal simulations presented in this work were performed with the MATLAB Reservoir Simulation Toolbox, MRST~\citep{krogstad2015,lie19,lie2021}. Specifically, we used the black-oil module, which is based on fully implicit solvers with automatic differentiation, and assigned properties of water to the oleic phase, such that the gaseous phase (CO$_2$ only) can dissolve in it. Vaporization of water into the gas phase and chemical reactions are not considered, since they are not primary controls on fluid migration for our operational setup and analysis time. 

In addition to structural and dissolution trapping, we also considered residual trapping~\citep{juanes2006} to be consistent with local measurements showing nonzero trapped gas saturation (sect.~\ref{sec:measurements}). This is achieved through hysteretic relative permeability curves for the nonwetting (gas) phase (see sect.~\ref{sec:models}). Our implementation in MRST follows ECLIPSE's technical description~\citep{schlumberger2014}, and~\cite{killough1976}'s model is used to compute the scanning curves~\citep[forthcoming]{saloMrst}. Physical diffusion was also included through the addition of a diffuse flux term with a scalar, constant coefficient in the computation of the total CO$_2$ flux~\citep{bear1972}. 

The simulator requires minute timesteps due to the buoyancy of CO$_2$ at atmospheric conditions and high sand permeabilities (Tab.~\ref{tab:measurements}). Linear solver time was reduced by means of AMGCL~\citep{demidov2018,lie19}, an external, pre-compiled linear solver. The greatest challenge was the convergence of the nonlinear solver, which required many iterations and timestep cuts. This is consistent with the experience reported from the groups working in the FluidFlower international benchmark study~\citep[this issue]{flemisch2023}.

Next, we describe the computational grids for experiments in tanks 1 and 2, PVT properties and boundary conditions. Petrophysical properties are model-specific and are detailed in sect.~\ref{sec:models}. 

\subsubsection{Computational grids}
\begin{figure}[h]%
\centering
\includegraphics[width=1\textwidth]{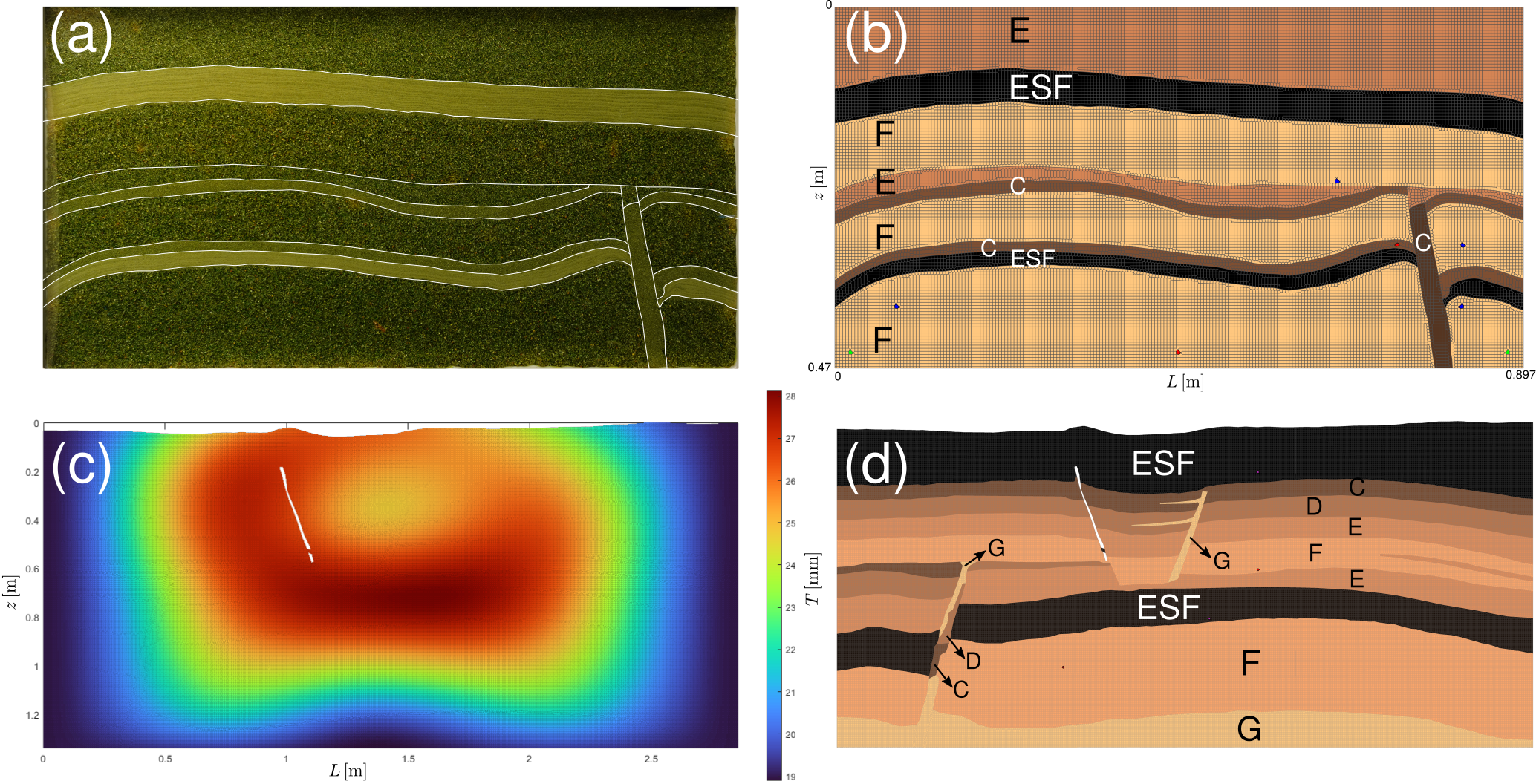}
\caption{Simulation grids overview. \textbf{a} front panel view of tank 1, where the layer contacts have been highlighted in white. \textbf{b} front view of simulation grid for experiments in tank 1, with lithologies indicated and colored based on petrophysical properties (see sect.~\ref{sec:models}). Location of injection wells is shown in red. \textbf{c} thickness map of simulation grid for experiments in tank 2. \textbf{d} front view of simulation grid for experiments in tank 2, with lithologies indicated and colored based on petrophysical properties.  Location of injection wells is shown in red.}
\label{fig:grid}
\end{figure}

A front panel image of the porous medium was used to obtain layer contact coordinates through a vector graphics software (Fig.~\ref{fig:grid}a). These contacts were then imported into MATLAB to generate the computational grids using the UPR module~\citep{berge2019,berge2021repo}(Fig.~\ref{fig:grid}b,d). The grids were generated in 2D and then extruded to 3D (using a single cell layer) to account for thickness and volume. Note that, in tank 1, where the porous medium has dimensions of $89.7\times47\times1.05$ cm, the thickness (space between the front and back panels) is constant (10.5 mm). Tank 2, which is significantly larger (porous medium dimensions $2.86\times1.3\times0.019$ m), has a thickness of 19 mm at the sides; however, it varies towards the middle due to forces exerted by the sand and water, to a maximum of 28 mm. A thickness map obtained after initial sand filling was used to generate our variable-thickness mesh via 2D interpolation (Fig.~\ref{fig:grid}c). Also, the top surface of the porous medium is not flat (height = $130 \pm 3$ cm).

Our composite Pebi grids~\citep{heinemann1991} have a Cartesian background and are refined around face constraints (contacts and faults) as well as cell constraints (injection wells)~\citep{berge2019,berge2021repo}. We generated multiple grids to test the finest grid we could afford to simulate experiment B1 in tank 2 with. Our grid has a cell size $h\approx 5$ mm and 151,402 cells (Fig.~\ref{fig:grid}d). The grid used for tank 1 has a similar cell size ($h\approx 4$ mm and 27,200 cells), which was chosen to reduce grid-size dependencies when applying our matched models to experiment B1. 

\subsubsection{PVT properties}
Consistent with experimental conditions, our simulations are conducted at atmospheric conditions ($T = 25$ C), where the CO$_2$ is in gaseous state. We employed a thermodynamic model based on the formulations by~\cite{dua03} and~\cite{spy03,spy05} to calculate the composition of each phase as a function of $p$, $T$. The implementation for a black-oil setup is described in~\cite{has08} and references therein. Given the boundary conditions (sect.~\ref{sec:bc}) and dimensions of our experimental porous media, pore pressure changes ($\Delta p$) are very small in our simulations (max $\Delta p \ll 1$ bar). Hence, the fluid properties remain similar to surface conditions, where the water and CO$_2$ have, respectively, a density of 997 and 1.78 kg/m$^3$, and a viscosity of 0.9 and 0.015 cP. The maximum concentration of CO$_2$ in water is $\approx 1.4$ kg/m$^3$. 


\subsubsection{Initial, boundary and operational conditions} \label{sec:bc}
Our porous media are fully saturated in water at the beginning of CO$_2$ injection. No-flow boundary conditions were applied everywhere except at the top boundary, which is at constant pressure and includes a fixed water table a few cm above the top of the porous medium. Injection is carried out via wells completed in a single cell at the corresponding coordinates. The diameter of injection wells is 1.8 mm in both tank 1 and tank 2, which operate at a constant flow rate (see sect.~\ref{sec:experiments}). The simulation injection schedule follows the experimental protocol, provided in Tab.~\ref{tab:experimentSchedule}. Note that injection rates in our simulations of experiment A1 and A2 were slightly adjusted during the calibration procedure, as explained in sect.~\ref{sec:calibration} and~\ref{sec:results}.


\subsection{Simulation models} \label{sec:models}
Three different models, I to III, are used throughout this study to evaluate the value of local data in predicting subsurface CO$_2$ migration. Each successive model was constructed based on access to an increasing level of local data, with model I having access to the least data and model III having access to the most data. The model-specific parameters are limited to the following:
\begin{itemize}
    \item Petrophysical properties (porosity, permeability, capillary pressure and relative permeability), which depend on available local data and are described in this section.
    \item The molecular diffusion coefficient ($D$). Models I-III were calibrated using the same value, $D = 10^{-9}$ m$^2$/s. Additionally, model III was also calibrated with $D = 3\times10^{-9}$ m$^2$/s.
    \item Injection rate. Experiments in tank 1 were conducted at a very low injection rate ($I_\text{R} = 2$ ml/min, see Tab.~\ref{tab:experimentSchedule}). Given that the mass flow controllers used in tank 1 may be inaccurate for this rate, the injection rate was also modeled as an uncertain parameter. Model calibration was achieved with $I_\text{R} \in [1.6,\, 1.8]$ ml/min for all three models.
\end{itemize}

All other model characteristics, including the grid and numerical discretization, remain unchanged. Below, we describe the starting petrophysical values for each of our three simulation models. Note that the experimental geometry in tank 1, used for matching, only contained sands ESF, C, E and F. Properties for sands D and G are also provided because they were required to simulate the experiment in tank 2 (Fig.~\ref{fig:experimentalSetup}). 

\subsubsection{Model I} \label{sec:modelI}
For this model, local petrophysical data was limited to a measure of the average grain size ($d$; see sect.~\ref{sec:measurements} and Tab.~\ref{tab:measurements}). Hence, petrophysical properties were estimated from published data in similar silica sands. Porosity was selected from data in~\cite{beard1973} and~\cite{smits2010} for moderately to well-sorted sands. Permeability was obtained from fitting a Kozeny-Carman model to data in~\cite{beard1973} and~\cite{trevisan2014}. The resulting equation has the form $k = \alpha d^2\phi^3$, where $\alpha$ equals 12,250.0 in our fit with $d$ in mm and $k$ in D. Obtained porosity and permeability values are provided in Table~\ref{tab:petroI}.

\begin{table}[h] 
\centering
\caption{Initial porosity and permeability for model I. See main text for estimation details.}
\begin{tabular}{@{}cccc@{}}
\toprule
Sand type & $d$ [mm] & $\phi$ [-] & $k$ [D] \\ \midrule
  ESF & 0.2 & 0.37 & 25 \\
  C & 0.66 & 0.38 &  290  \\
  D & 1.05 & 0.40 &  930  \\
  E & 1.45 & 0.39 &  1530  \\
  F & 1.77 & 0.39 &  2280  \\
  G & 2.51 & 0.42 &  5720  \\ \bottomrule
\end{tabular}
\label{tab:petroI}
\end{table} 

\begin{figure}[h]%
\centering
\includegraphics[width=1\textwidth]{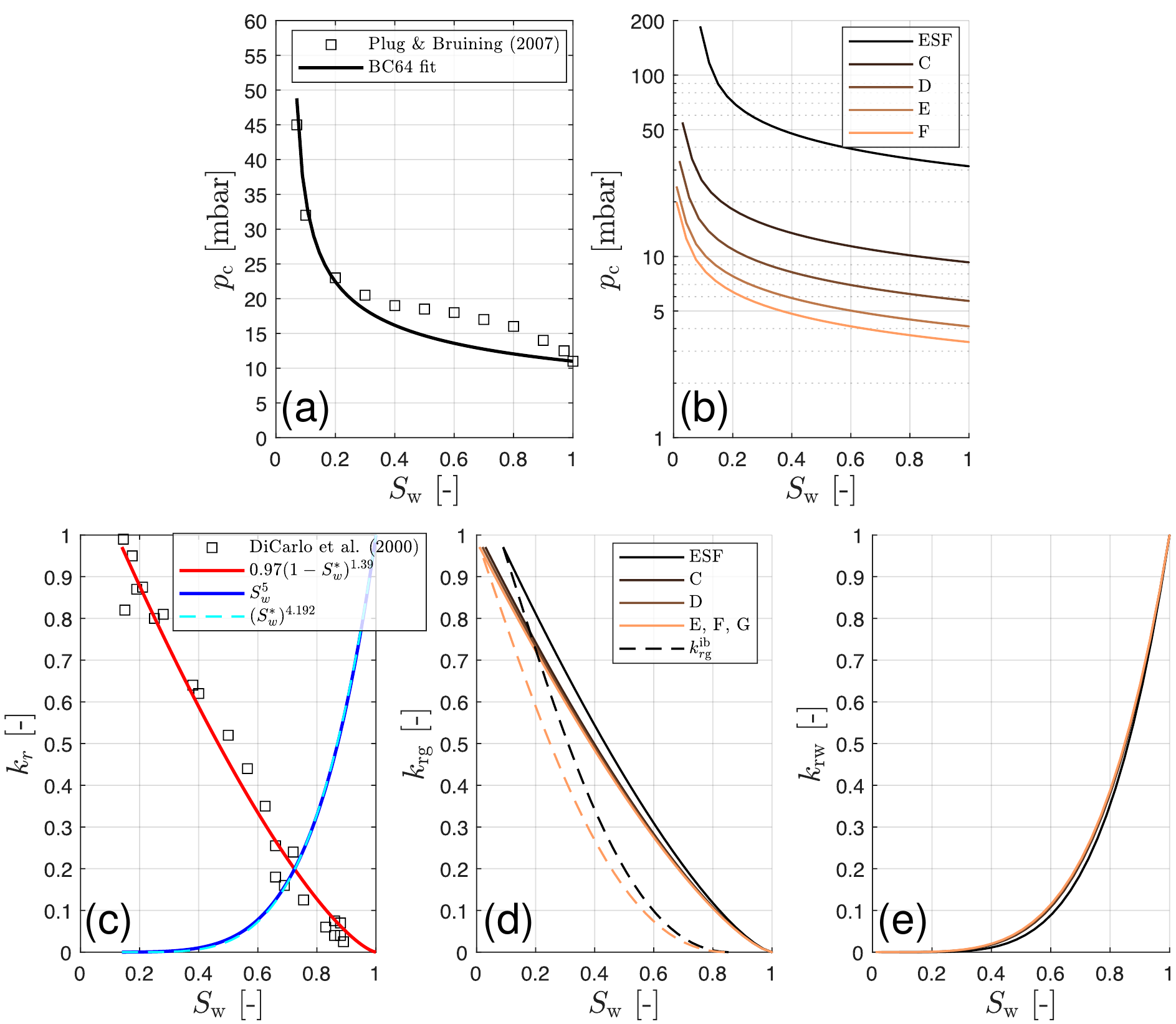}
\caption{Multiphase flow properties for model I. \textbf{a} Capillary pressure measurements and reference curve using a~\cite{brooks1964} function. \textbf{b} Initial capillary pressure curves, computed from the reference curve using Leverett scaling (see main text). \textbf{c} Relative permeability data (squares and $S_\text{w}^5$ model) and our fitted Corey model. \textbf{d,e} Relative permeability of gas and water, respectively. The drainage curve is shown as a solid line, while the bounding imbibition curve is shown for sands ESF and G as a discontinuous line. No relative permeability hysteresis was considered for the water phase.}
\label{fig:krpc}
\end{figure}

Capillary pressure curves were computed as described below: 
\begin{enumerate}
    \item Capillary pressure measurements in a similar system were obtained from the literature. In this case,~\cite{plug2007} measured capillary pressure curves on the unconsolidated quartz sand-CO$_2$-distilled water system at atmospheric conditions. We used their measurements on sand packs with an average particle size between 0.36 and 0.41 mm, which are closest to the C sand in our experiments (Fig.~\ref{fig:krpc}a).
    \item A~\cite{brooks1964} model of the form $p_\text{c} = p_\text{e}(S_\text{w}^*)^{-\frac{1}{\lambda}}$ was fitted to these data, where $p_\text{e}$ is the nonwetting phase entry pressure at $S_\text{w} = 1$, $\lambda = 2.6$ and $S_\text{w}^* = \frac{S_\text{w} - S_{\text{wc}}}{1 - S_{\text{wc}}}$ is the normalized water saturation with irreducible or connate water saturation $S_\text{wc}$. This fit led to our reference curve, $p_\text{cr}$ (Fig.~\ref{fig:krpc}a). 
    \item The capillary pressure depends on the pore structure of each material, such that sands with different grain sizes require different $p_\text{c}$ curves. The capillary pressure variation can be modeled by means of the dimensionless $J$-function proposed by Leverett~\citep{leverett1941,saadatpoor2010}: $J(S_\text{w}) = \frac{p_\text{c}}{\sigma\cos\theta}\sqrt{\frac{k}{\phi}}$, where $\sigma$ is the surface tension and $\theta$ the contact angle. Assuming the same wettability and surface tension for different sand regions, and the same shape of the $p_\text{c}$ curve, the capillary pressure for any given sand ($p_\text{cs}$) can be obtained from the reference curve as $p_\text{cs}(S_\text{w}) = p_\text{cr}(S_\text{w})\sqrt{\frac{k_\text{r} \phi_\text{s}}{k_\text{s}\phi_\text{r}}}$ (Fig.~\ref{fig:krpc}b).
\end{enumerate}

Drainage relative permeabilities were obtained from CO$_2$-water measurements by~\cite{dicarlo2000}, who used water-wet sandpacks with 0.25 mm grain size. Specifically, we used the data reported in their Fig. 4 and 5, and fitted Corey-type functions~\citep{corey1954,brooks1964} of the form $k_\text{rw} = (S_\text{w}^*)^a$ and $k_\text{rg} = c(1-S_\text{w}^*)^b$ (Fig.~\ref{fig:krpc}c). The fitted exponents $a$ and $b$ are 4.2 and 1.4, respectively, while $c$ is 0.97. We assumed that the difference in relative permeability of different sands is the result of different irreducible water saturation only (see Fig.~\ref{fig:krpc}d,e). For each of our sands, $S_\text{wc}$ was obtained from~\cite{timur1968} as $S_\text{wc} = 0.01\times3.5\frac{\phi^{1.26}}{k^{0.35}}-1$, where $\phi$ is in percent and $k$ in mD. This model was used to compute $S_\text{wc}$ for both the $p_\text{c}$ and $k_\text{r}$ curves.

In CO$_2$ storage, secondary imbibition occurs where the water displaces buoyant gas at the trailing edge of the CO$_2$ plume, disconnecting part of the CO$_2$ body into blobs and ganglia and rendering them immobile~\citep[and references therein]{juanes2006}. This means that the maximum water saturation that can be achieved during imbibition equals 1 - $S_\text{gt}$ (the trapped gas saturation). Here, we used measurements in sandpacks from~\cite{pentland2010} to determine $S_\text{gt}$. In particular, we fitted~\cite{land1968}'s model with the form $S_\text{gt}^* = \frac{S_\text{gi}^*}{1+CS_\text{gi}^*}$, where $S_\text{g}^* = \frac{S_\text{g}}{1-S_\text{wc}} = 1 - S_\text{w}^*$, $S_\text{gi}$ is the gas saturation at flow reversal, and $C$ is Land's trapping coefficient with a value of 5.2 in our fit. Although~\cite{pentland2010} report that the best fit is achieved with the~\cite{aissaoui1983} and~\cite{spiteri2008} models (cf. their Fig. 5), Land's model was chosen here given that most relative permeability hysteresis models build on this one (see next paragraph).

Nonwetting phase trapping contributes to irreversibility of the relative permeability and capillary pressure curves (hysteresis). Here, we accounted for this mechanism in the gas relative permeability due to its importance in subsurface CO$_2$ migration~\citep[and references therein]{juanes2006}. In particular, we used~\cite{land1968}'s model to compute the bounding imbibition curve (see Fig.~\ref{fig:krpc}d), where $S_\text{gt}$ is obtained as described above, and~\cite{killough1976}'s model to characterize the scanning curves. In Killough's model, the scanning curves are reversible, such that the relative permeability at $S_\text{g} < S_\text{gi}$ no longer depends on the displacement type~\citep[forthcoming]{saloMrst}. 

\subsubsection{Model II} \label{sec:modelII}
This model had access to local, ex-situ measurements of single-phase petrophysical properties, i.e., porosity and intrinsic permeability (see sect.~\ref{sec:measurements} and Tab.~\ref{tab:measurements}). Comparing with Tab.~\ref{tab:petroI}, it can be seen that our estimation for model I above was correct to the order of magnitude, but resulted in smaller values: porosity $\in [85,\, 93]\%$ and permeability $\in [53,\, 84]\%$ of the local measurements.

\begin{figure}[h]%
\centering
\includegraphics[width=1\textwidth]{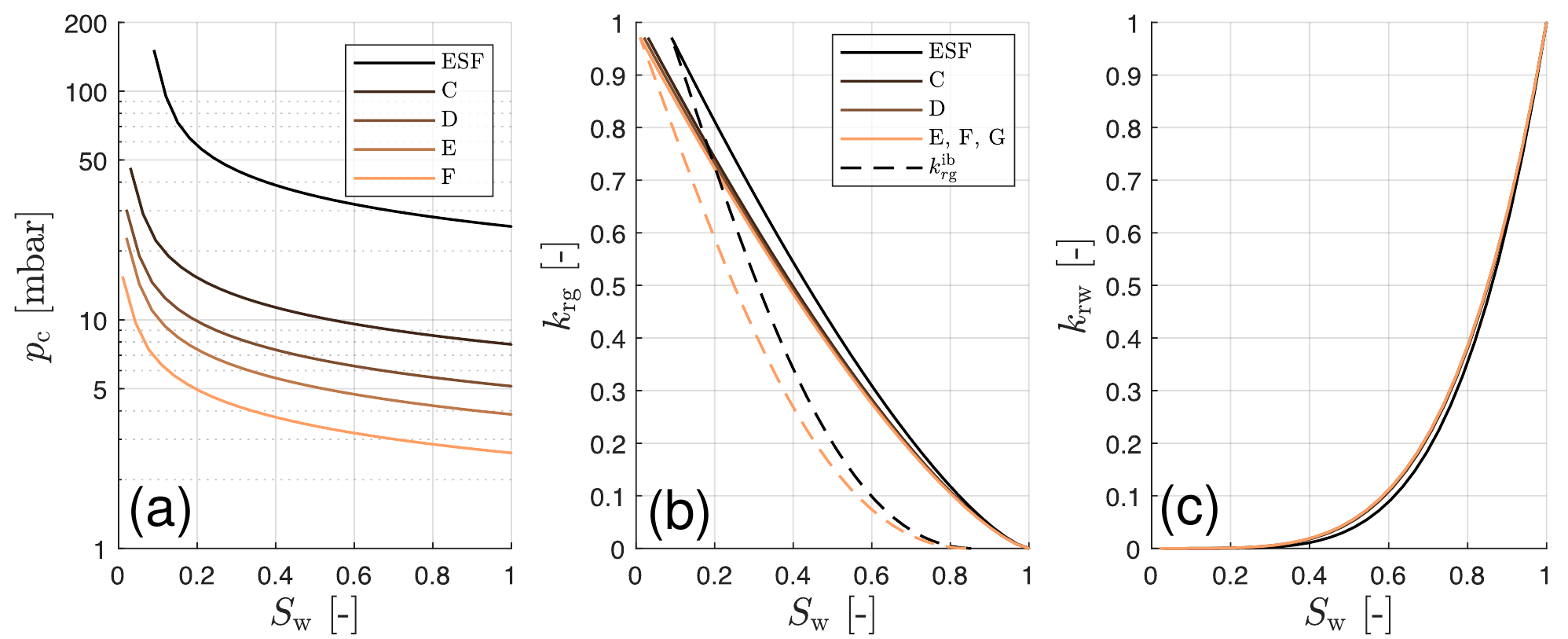}
\caption{Multiphase flow properties for model II. \textbf{b} Initial capillary pressure curves, computed from the reference curve using Leverett scaling (see main text).  \textbf{b,c} Relative permeability of gas and water, respectively. The drainage curve is solid, while the bounding imbibition curve is shown for sands ESF and G as a discontinuous line. No relative permeability hysteresis was considered for the water phase.}
\label{fig:krpc2}
\end{figure}

Capillary pressures and relative permeabilities were obtained using the same procedure described above for model I. The slight differences with respect to the curves shown in Fig.~\ref{fig:krpc}b,d,e come from the porosity and permeability values used in the Leverett scaling and to determine $S_\text{wc}$, which were taken from Tab.~\ref{tab:measurements} instead. The obtained curves for model II are provided in Fig~\ref{fig:krpc2}.

\subsubsection{Model III}
\begin{figure}[h]%
\centering
\includegraphics[width=1\textwidth]{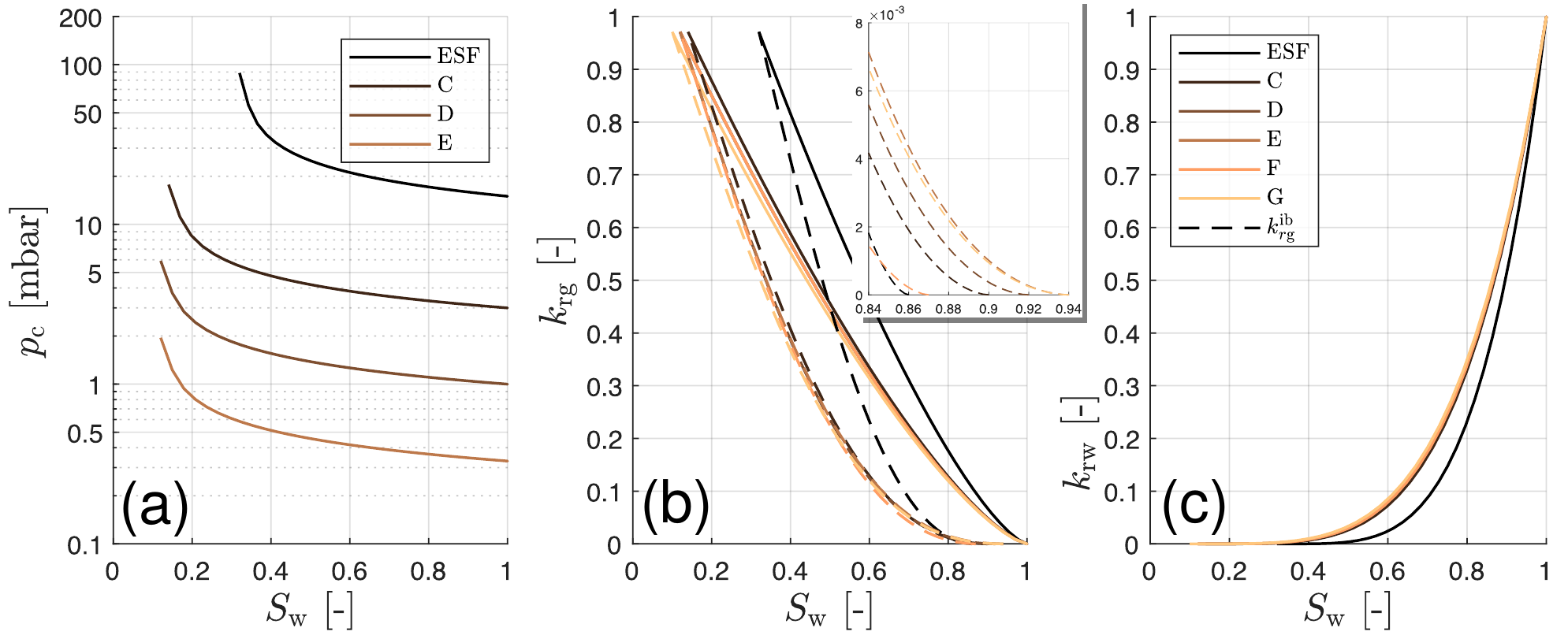}
\caption{Multiphase flow properties for model III. \textbf{b} Initial capillary pressure curves, computed according to the entry pressure determined experimentally (see sect.~\ref{sec:measurements}). \textbf{b,c} Relative permeability of gas and water, respectively, according to the endpoints determined experimentally (sect.~\ref{sec:measurements}). The drainage curves are solid, while the bounding imbibition curves are shown as a discontinuous line. The inset in \textbf{b} is a zoom view around the trapped gas saturation. No relative permeability hysteresis was considered for the water phase.}
\label{fig:krpc3}
\end{figure}

This model was allowed access to all local, ex-situ measurements (see Tab.~\ref{tab:measurements}). Initial porosity and permeability remain unchanged with respect to model II. Capillary pressure curves were obtained by scaling the reference curve described in sect.~\ref{sec:modelI} and shown in Fig.~\ref{fig:krpc}a using the measured entry pressure (sect.~\ref{sec:measurements}). The scaling followed the model $p_\text{cs}(S_\text{w}) = p_\text{cr}(S_\text{w})\frac{p_\text{e}}{p_\text{er}}$, where  $p_\text{e}$ is the measured entry pressure for each sand, and $p_\text{er}$ is the reference curve entry pressure. The obtained curves are shown in Fig.~\ref{fig:krpc3}a.

Relative permeabilities were computed following the same procedure described for model I above. In this case, however, each sand type was assigned the measured $S_\text{wc}$ and $S_\text{gt}$ values (see Tab.~\ref{tab:measurements}). This led to differences in both the drainage and imbibition curves, as shown in Fig.~\ref{fig:krpc3}. 

\subsection{Model calibration} \label{sec:calibration}
\begin{figure}[h]%
\centering
\includegraphics[width=1\textwidth]{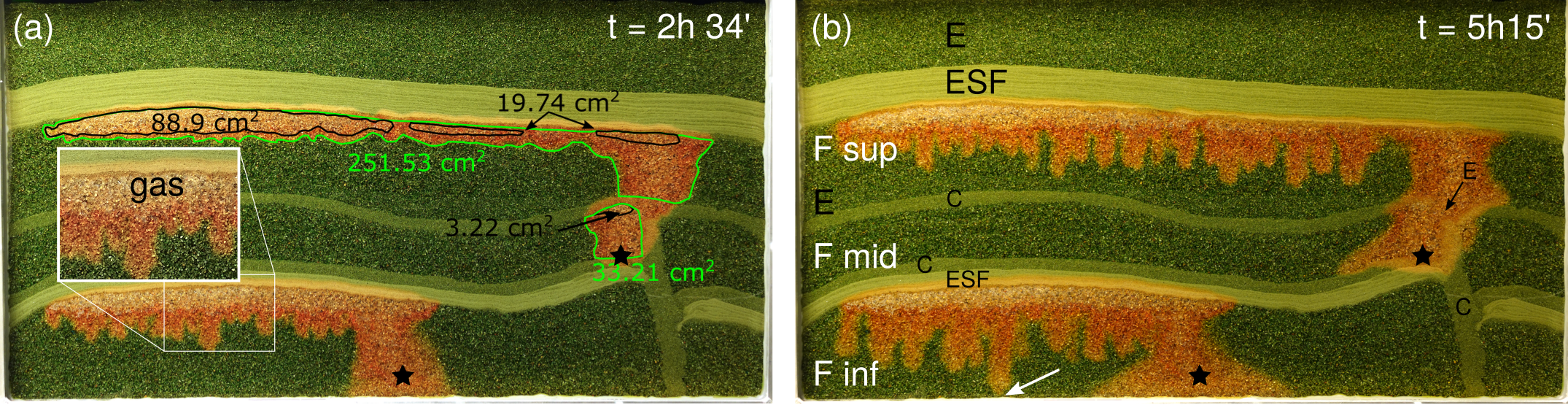}
\caption{Front panel view of tank 1, showing quantities and times for history matching of numerical models to experiment A1. \textbf{a} shows areas with gaseous CO$_2$ (free-phase) and dyed water with dissolved CO$_2$ at the end of injection. Location of injection ports is shown with a star. \textbf{b} shows the time and location where the first finger touches the bottom of the tank (white arrow), as well as the different lithological units. Note the three F reservoirs labeled `inf', `mid' and `sup', mentioned in the text and other figures.}
\label{fig:experimentalvolume}
\end{figure}

The mismatch between results obtained with each simulation model (I to III) and the validation experiment in tank 1 (A1, see sect.~\ref{sec:experimentSchedule}) is quantitatively assessed by comparing the following quantities (see Fig.~\ref{fig:experimentalvolume}):
\begin{enumerate}
    \item At $t = 55$ min  (end of injection in port $I_1$): Areas occupied by free-phase CO$_2$, and dyed water with dissolved CO$_2$ in the bottom F reservoir. 
    \item At $t = 154$ min (end of injection in  port $I_2$): Areas occupied by free-phase CO$_2$, and dyed water with dissolved CO$_2$, in the middle and top F reservoirs.
    \item Time at which the first finger touches the tank bottom.
    \item Time at which the first finger (sinking from the top F reservoir) touches the middle C sand.
\end{enumerate}
The experimental values for points 1-2 were obtained by computing areas from time-lapse images using a vector-graphics software. Careful visual inspection of color-enhanced images was used to distinguish between free-phase CO$_2$ (white) and dyed water with dissolved CO$_2$ (yellowish orange to red), and to identify the times for points 3-4 above. In the simulation models, the threshold gas saturation and CO$_2$ concentration in water used to compute areas were $S_\text{g} > 10^{-3}$ and $C_{\text{CO}_2} > 15\%(C_{\text{CO}_2}^\text{max})\approx0.21$ [kg/m$^3$], respectively. The $C$ value was chosen after a shape comparison of the region with dissolved CO$_2$. A smaller value of $C_{\text{CO}_2} > 0.05$ [kg/m$^3$] was selected to determine finger times for points 3 and 4 above. Fig.~\ref{fig:experimentalvolume} shows an overview of the experimental values for points 2 and 3, while Fig.~\ref{fig:match} in Sect.~\ref{sec:hm} shows the full comparison with the 
history-matched/calibrated simulation models.

The experiment was conducted first. Afterwards, the process consisted of running simulation models I to III, in parallel, starting with the petrophysical properties described in sect.~\ref{sec:models}. Given the number of uncertain variables (four petrophysical properties for each lithological unit, the diffusion coefficient and the injection rate) and the time required to complete a single simulation, a manual history matching method was employed. At the end of each run, quantities 1-4 above were compared and one or more properties were manually changed based on observed mismatch and domain knowledge. During the first few runs, only quantities 1 and 2 above were compared. After obtaining a satisfactory areal match, petrophysical properties were further adjusted to match quantities 3 and 4. 

\section{Results} \label{sec:results}
In sect.~\ref{sec:run1}, we present the results of the first simulation of experiment A1 with each model and property values detailed in sect.~\ref{sec:models}. Then, we detail the calibration of simulation models using experiment A1, and assess the value of local data to history-match CO$_2$ storage simulation models (sect.~\ref{sec:hm}). Finally, we apply these matched models to experiment A2, analog for a longer injection in the same geology (sect.~\ref{sec:a2}), and to experiment B1, analog for a larger-scale injection in a different geologic setting (sect.~\ref{sec:benchmark}). We use simulations of experiments A2 and B1 to assess the predictive ability of simulation models in different conditions.

\subsection{Initial model results} \label{sec:run1}
\begin{figure}[h]%
\centering
\includegraphics[width=1\textwidth]{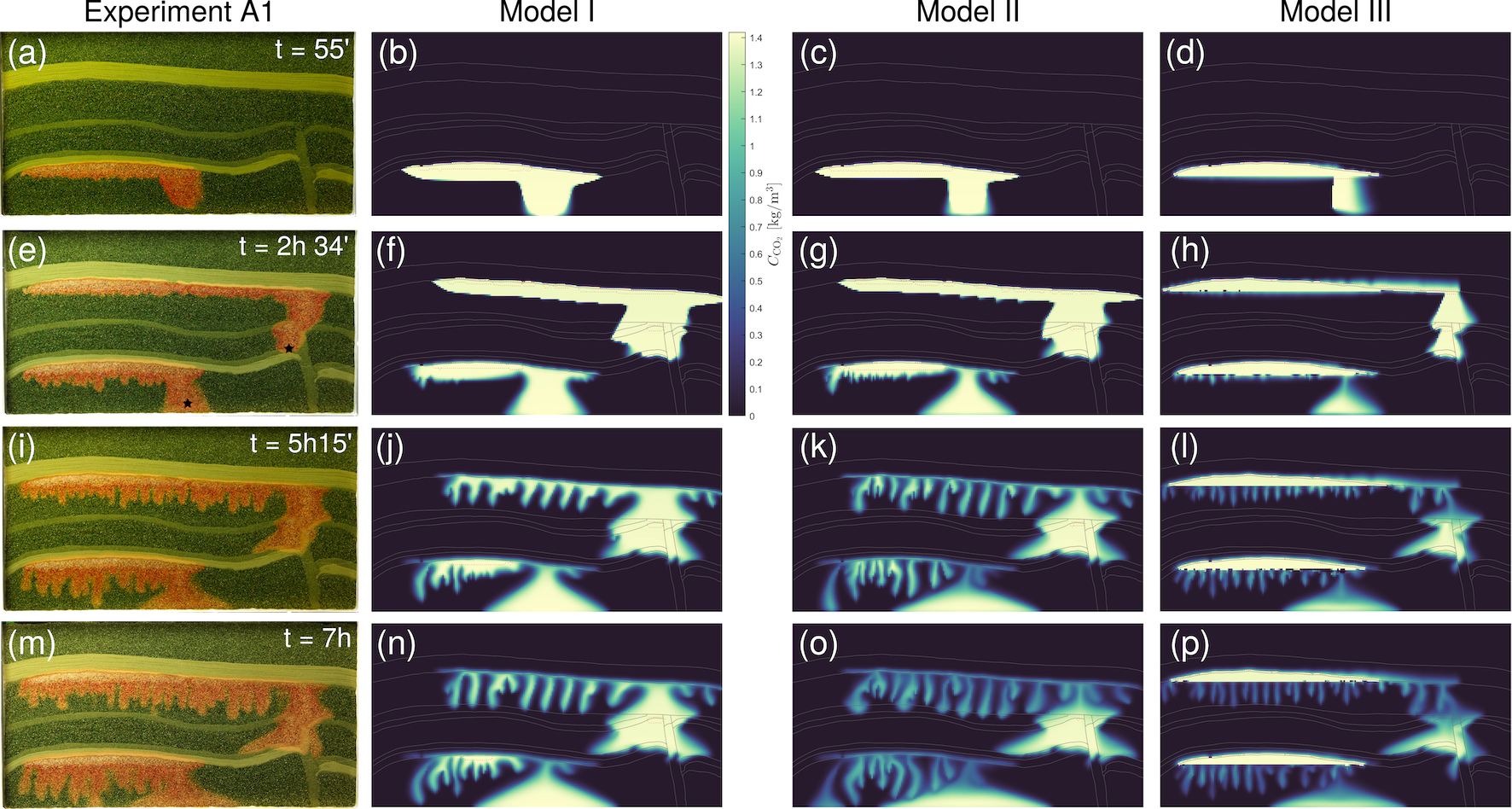}
\caption{Comparison between experiment A1 in tank 1 (left column) and first run simulation results with models I-III. Color map in simulation plots refers to CO$_2$ concentration in water, according to color bar. The dotted red contours in simulation plots indicate $S_\text{g} = 10^{-3}$. \textbf{a-d}: end of injection in port 1. \textbf{e-h}: end of injection in port 2. \textbf{i-l}: time at which the first finger touches the tank bottom. \textbf{m-p}: time at which the first finger touches the middle C sand.}
\label{fig:run1}
\end{figure}

Fig.~\ref{fig:run1} shows the comparison between experiment A1 and the first run with each model, at times indicated in sect.~\ref{sec:calibration}. Numerous differences are evident between the experiment and models I and II, while model III is much closer to the experiment. In particular, models I and II overestimate the extent of CO$_2$-rich brine and underestimate the amount of gaseous CO$_2$ in all F reservoirs (refer to Fig.~\ref{fig:experimentalvolume} for location). Model III approximates much better the areal extent of gaseous CO$_2$ in all regions, as well as the CO$_2$-rich brine in the middle and upper F reservoirs. Model II provides the closest finger migration times (points 3 and 4 in sect.~\ref{sec:calibration}), although this was not evaluated in the first run, as discussed below.

Petrophysical properties for models I and II were obtained from references in sect.~\ref{sec:models}, which also used silica sands with similar grain sizes. However, despite the relatively homogeneous nature of our quartz sands, model III performed significantly better. This result stems from natural sand variability and highlights the difficulty in establishing general, representative elementary volume-scale properties for porous media~\citep[see, for instance,][for a discussion on intrinsic permeability]{hommel2018,schulz2019}. Additionally, results in Fig.~\ref{fig:run1} highlight the need for conducting sand/rock-specific measurements, even in the case of well-sorted, homogeneous sediments.

\subsection{Manual history matching and value of local data} \label{sec:hm}
\begin{figure}[h]%
\centering
\includegraphics[width=1\textwidth]{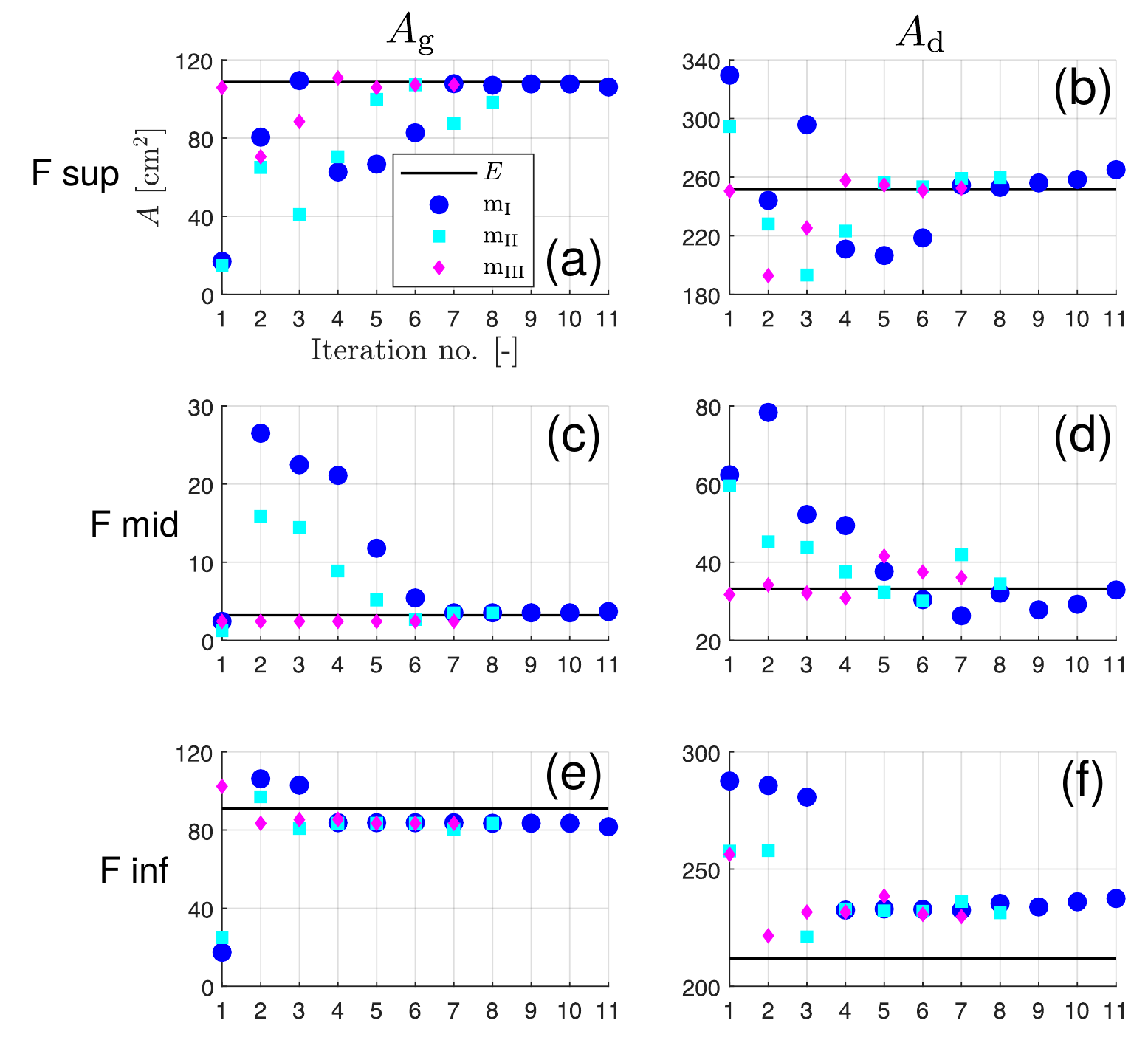}
\caption{Convergence of areas occupied by free gas ($A_\text{g}$, left column) and water with dissolved CO$_2$ ($A_\text{d}$, right column), during the calibration of models I-III with experiment A1. $A_\text{d}$ includes area with gaseous CO$_2$ (see Fig.~\ref{fig:experimentalvolume}). Each iteration represents a new simulation run, and the experimental reference ($E$) is shown as a black line. Refer to Fig.~\ref{fig:experimentalvolume} for region location, and to sect.~\ref{sec:calibration} for calibration procedure. \textbf{a,b}: upper F sand. \textbf{c,d}: middle F sand. \textbf{e,f}: lower F sand.}
\label{fig:AC02_areas}
\end{figure}

\begin{figure}[h]%
\centering
\includegraphics[width=0.5\textwidth]{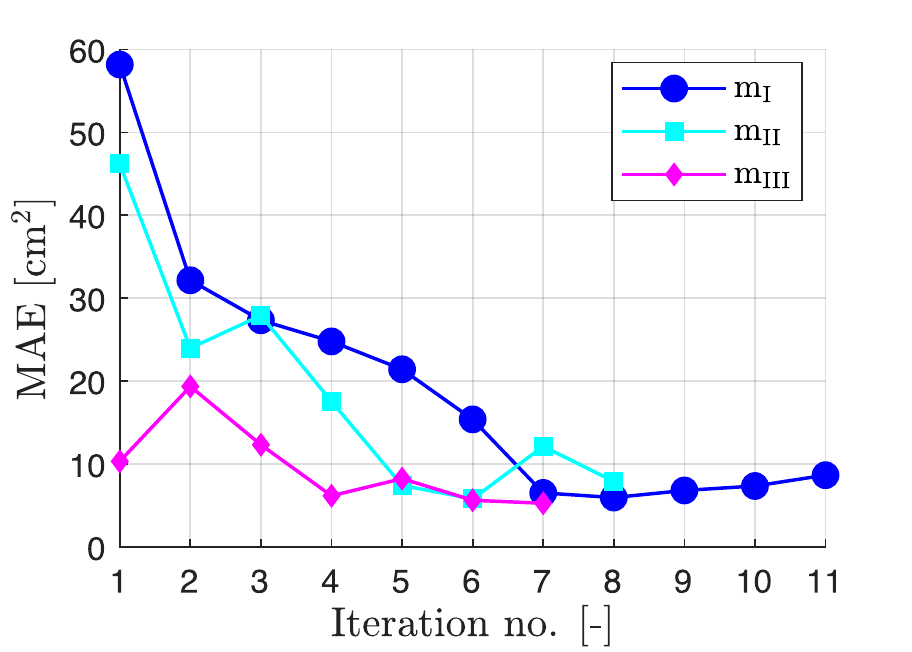}
\caption{Convergence of mean absolute error over the six areal quantities measured during the calibration process. The error is computed with respect to experimental values. See Fig.~\ref{fig:AC02_areas} for areas measured, and refer to sect.~\ref{sec:calibration} for calibration procedure.}
\label{fig:AC02_mae}
\end{figure}

Fig.~\ref{fig:AC02_areas} shows convergence of areas occupied by free gas ($A_\text{g}$) and water with dissolved CO$_2$ ($A_\text{d}$), according to sect.~\ref{sec:calibration}. Each iteration corresponds to a successive model with updated parameters, and the different F sand regions evaluated in each panel (a) to (f) are provided in Fig.~\ref{fig:experimentalvolume}. With the exception of $A_\text{d}$ in the upper compartment, model III performs well since the beginning, and all areas were satisfactorily matched after four iterations. Conversely, model I and II were significantly off the experimental reference during the first few iterations. Model II, however, performed well after five iterations, while model I required seven iterations to give satisfactory areal estimates. The mean absolute error (MAE) over the six areal quantities presented in Fig.~\ref{fig:AC02_areas} is evaluated in Fig.~\ref{fig:AC02_mae}, where it can be seen that, while all models are accurate towards the end (MAE $\in [5-10]$ cm$^2$), that required a six-fold improvement in models I and II, but only two-fold in model III. As mentioned in sect.~\ref{sec:calibration}, $C_{\text{CO}_2} > 15\%(C_{\text{CO}_2}^\text{max})\approx0.21$ [kg/m$^3$] was used as threshold to determine areas. While the absolute values and error would change with a different $C_{\text{CO}_2}$ threshold, we checked that the relative performance of our calibrated models does not with both $C_{\text{CO}_2} > 0.01$ and 0.1 [kg/m$^3$].

\begin{figure}[h]%
\centering
\includegraphics[width=0.8\textwidth]{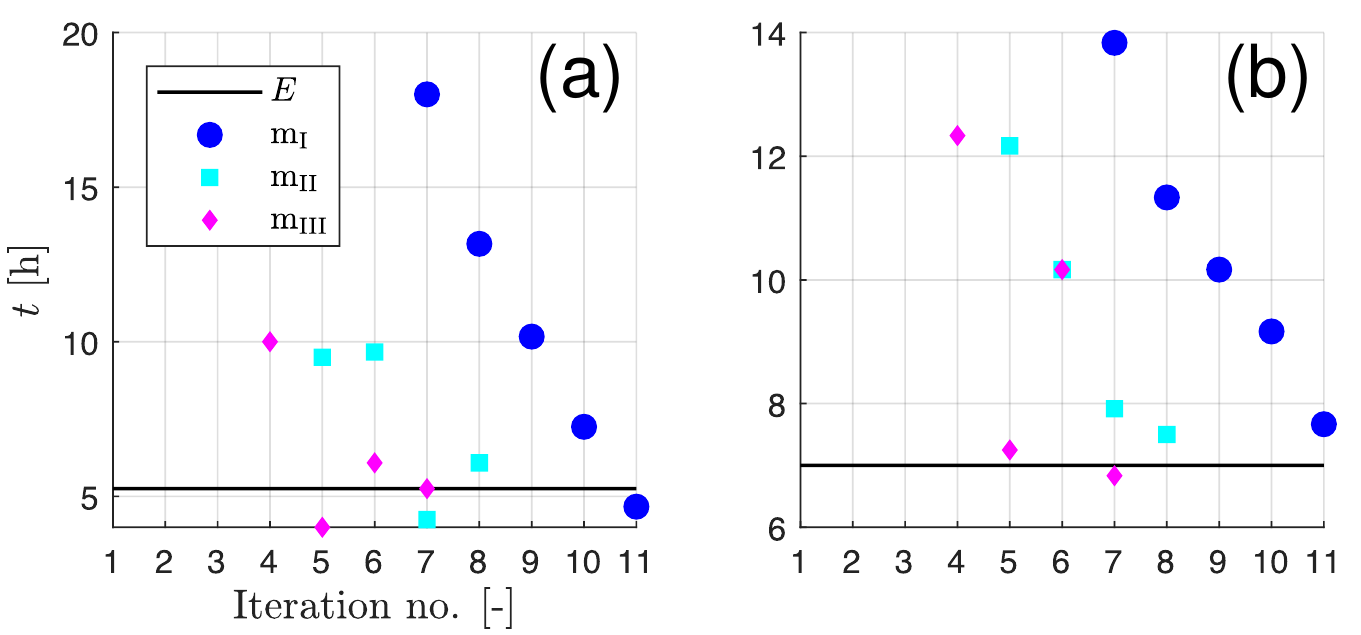}
\caption{Convergence of times at which the first finger touches the bottom of the rig (\textbf{a}) and the middle C sand (\textbf{b}), during the calibration of models I-III with experiment A1. Refer to sect.~\ref{sec:calibration} for calibration procedure.}
\label{fig:AC02_tfinger}
\end{figure}

Convergence of quantities 3 and 4 in sect.~\ref{sec:calibration}, the times at which the first finger touches the rig bottom and the middle C sand, respectively, are provided in Fig.~\ref{fig:AC02_tfinger}. These times were only evaluated after a satisfactory areal match for quantities in Fig.~\ref{fig:AC02_areas} was achieved. Therefore, areas no longer change much in the last few iterations in Fig.~\ref{fig:AC02_areas}. In Fig.~\ref{fig:AC02_tfinger}, it can be seen that model II and III, which incorporated local intrinsic permeability measurements, were significantly closer to our experimental reference than model I. Initially, however, we observed that sinking of gravity fingers in the experiment was faster than our model values by a factor of $\approx 2$. A satisfactory match of all quantities evaluated was achieved after 11, 8, and 7 iterations for models I-III, respectively.

Overall, we find that model III, with access to local single-phase and multiphase flow properties, is closer to the experimental reference from the start. Model I started farthest, and required significantly more effort for calibration. After the calibration process, all models achieve a similar, accurate estimate of the evaluated quantities (Fig.~\ref{fig:AC02_mae}). 

The calibration shown in Fig.~\ref{fig:AC02_areas},~\ref{fig:AC02_mae},~\ref{fig:AC02_tfinger} employs $D = 10^{-9}$ m$^2$/s in all models. Injection rates ($I_\text{R}$) started at 2.0 ml/min for all three models, and were 1.6 ml/min, 1.8 ml/min and 1.75 ml/min, respectively, at the end of the calibration. $I_\text{R}$ is slightly different because the goal was to obtain the best match with each model, considering $I_\text{R}$ to be an uncertain variable. In sect.~\ref{sec:predictions} below, the same $I_\text{R}$ is used to make predictions with all three models. 

Tab.~\ref{tab:matchedProperties} compares the starting and final (matched) key petrophysical variables for each model. The models were successfully calibrated by adjusting intrinsic permeability and the capillary pressure curves (same shape, but scaled to higher or lower $p_\text{e}$) only. It was found that CO$_2$ migration was most sensitive to the properties of the F sand, were most of the CO$_2$ migration occurs, as well as the ESF seal, which structurally traps the CO$_2$ plume. In our matched models, $p_\text{e}$ of ESF is about twice the measured value; this was required because the minimum saturation at which we can define $p_\text{e}$ and ensure convergence is $S_\text{g} \approx 10^{-4}$. Reality, however, is closer to a jump in $p_\text{c}$ from 0 to $p_\text{e}$ at an infinitesimally small $S_\text{g}$. Additionally, we found that our simulations performed better when using different values for the C and F sands in different model regions. In the case of the C sand, the explanation lies in the fault construction process, which may reduce porosity with respect to ``natural" sedimentation of stratigraphic layers~\citep[this issue]{haugen2023}. The increase in F sand permeability was required to match finger migration times, and is possibly compensating the absence of mechanical dispersion in the simulations. This is discussed in sect.~\ref{sec:discussion}. Our calibrated values are within the same order of magnitude of the ex-situ measurements (Tab.~\ref{tab:matchedProperties}) and history-matched values for the porous medium in tank 2~\citep[forthcoming]{landa2023}.

\begin{table}
\centering
\caption{Petrophysical properties for used quartz sands in experiment A1. Methodology for local measurements is provided by~\cite{haugen2023}, while starting property modeling is described in sect.~\ref{sec:models}. For each sand, measured (first row), initial (superscript i) and final (superscript f) values for each of our models is shown. For sand C, the second permeability value refers to the fault, if different from the rest. For sand F, the second permeability value refers to the middle F layer, if different from the rest. For model III, where property values are different, m$_{\text{III},1}$ refers to the calibration with $D = 10^{-9}$ m$^2$/s and m$_{\text{III},3}$ refers to $D = 3\times10^{-9}$ m$^2$/s.}
{\renewcommand{\arraystretch}{.8}
\resizebox{!}{.29\paperheight}{%
\begin{tabular}{cccccc}
\toprule
Sand type / model & $\phi$ [-] & $k$ [D] & $p_\text{e}$ [mbar] & $S_\text{wc}$ [-] & $S_\text{gt}$ [-]\\ \midrule
  ESF & 0.435 & 44 & 15 & 0.32 & 0.14\\
  m$_\text{I}^\text{i}$ & 0.37 & 25 & 31.4 & 0.09 & 0.1468 \\
  m$_\text{I}^\text{f}$ & 0.37 & 6 & 31.4 & 0.09 & 0.1468 \\
  m$_\text{II}^\text{i}$ & 0.435 & 44 & 25.6 & 0.09 & 0.1468 \\
  m$_\text{II}^\text{f}$ & 0.435 & 44 & 25.6 & 0.09 & 0.1468 \\
  m$_\text{III}^\text{i}$ & 0.435 & 44 & 15 & 0.32 & 0.14 \\
  m$_\text{III}^\text{f}$ & 0.435 & 15 & 30 & 0.32 & 0.14 \\ \midrule
  C &  0.435 & 473 & 3 & 0.14 & 0.1 \\
  m$_\text{I}^\text{i}$ & 0.38 & 293 & 9.3 & 0.03 & 0.1565 \\
  m$_\text{I}^\text{f}$ & 0.38 & 293, 27 & 4.6 & 0.03 & 0.1565 \\
  m$_\text{II}^\text{i}$ & 0.435 & 473 & 7.8 & 0.03 & 0.1565 \\
  m$_\text{II}^\text{f}$ & 0.435 & 473, 158 & 2.6 & 0.03 & 0.1565 \\
  m$_\text{III}^\text{i}$ & 0.435 & 473 & 3 & 0.14 & 0.1 \\
  m$_\text{III}^\text{f}$ & 0.435 & 473, 118 & 4.5 & 0.14 & 0.1 \\ \midrule
  E &  0.45 & 2005 & - & 0.12 & 0.06 \\
  m$_\text{I}^\text{i}$ & 0.39 & 1528 & 4.1 & 0.01 & 0.16 \\
  m$_\text{I}^\text{f}$ & 0.39 & 1528 & 0.5 & 0.01 & 0.16 \\
  m$_\text{II}^\text{i}$ & 0.45 & 2005 & 3.86 & 0.01 & 0.16 \\
  m$_\text{II}^\text{f}$ & 0.45 & 3008 & 0.58 & 0.01 & 0.16 \\
  m$_\text{III}^\text{i}$ & 0.45 & 2005 & 0.33 & 0.12 & 0.06 \\
  m$_{\text{III},1}^\text{f}$ & 0.45 & 2406 & 0.33 & 0.12 & 0.06 \\ 
  m$_{\text{III},3}^\text{f}$ & 0.45 & 3208 & 0.33 & 0.12 & 0.06 \\
  \midrule
  F &  0.44 & 4259 & - & 0.12 & 0.13 \\ 
  m$_\text{I}^\text{i}$ & 0.39 & 2277 & 3.3 & 0.01 & 0.16 \\
  m$_\text{I}^\text{f}$ & 0.39 & 6540, 2907 & 0 & 0.01 & 0.16 \\
  m$_\text{II}^\text{i}$ & 0.44 & 4259 & 2.62 & 0.01 & 0.16 \\
  m$_\text{II}^\text{f}$ & 0.44 & 6814, 4259 & 0 & 0.01 & 0.16 \\
  m$_\text{III}^\text{i}$ & 0.44 & 4259 & 0 & 0.12 & 0.13 \\
  m$_{\text{III},1}^\text{f}$ & 0.44 & 7240, 4685 & 0 & 0.12 & 0.13 \\
  m$_{\text{III},3}^\text{f}$ & 0.44 & 9796, 4259 & 0 & 0.12 & 0.13 \\ \bottomrule
\end{tabular}}}
\label{tab:matchedProperties}
\end{table} 

\begin{figure}[h]%
\centering
\includegraphics[width=1\textwidth]{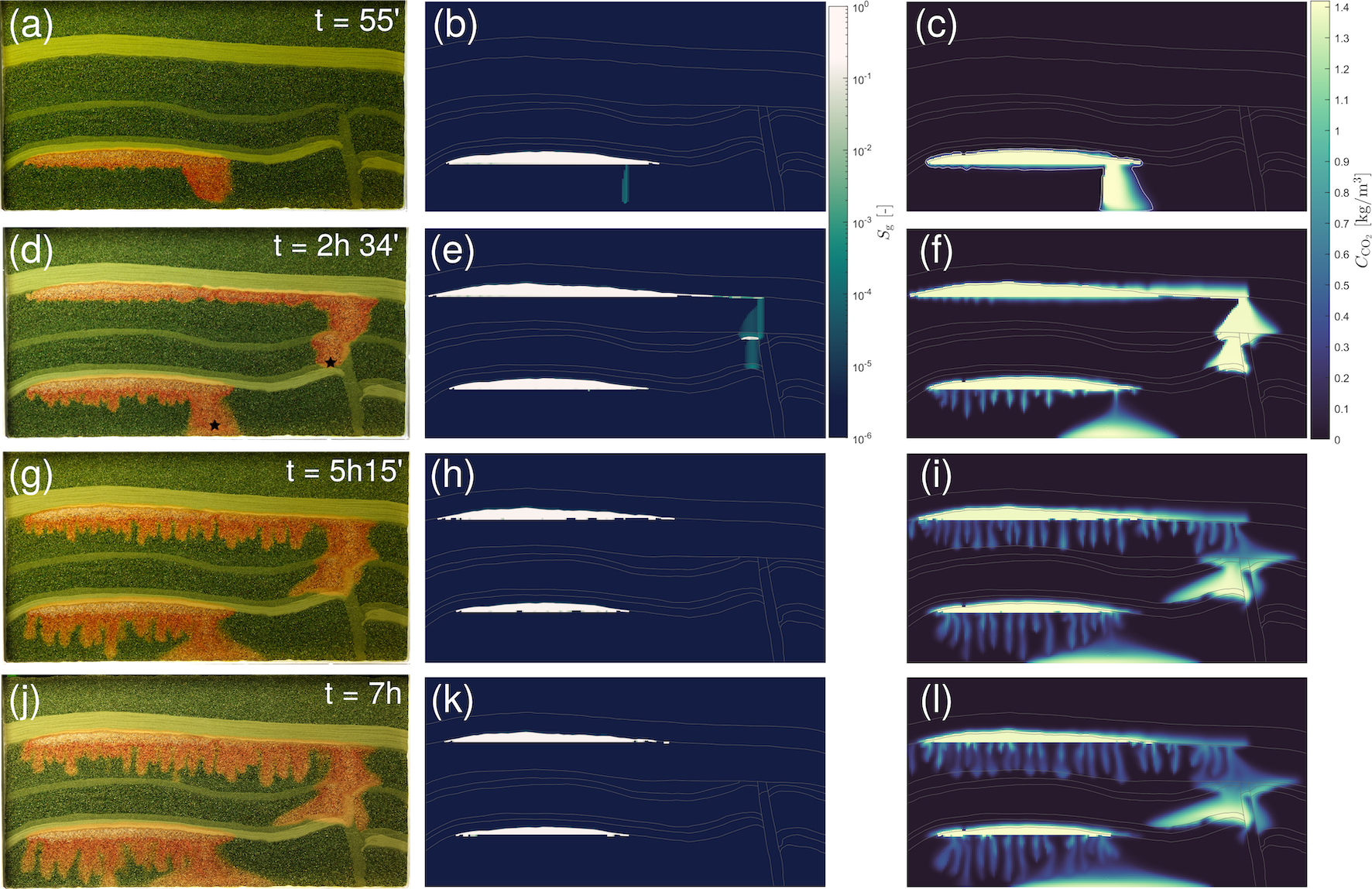}
\caption{Comparison between experiment A1 in tank 1 (left column) and simulation results with model III after calibration (gas saturation shown in middle column, and CO$_2$ concentration shown in right column). Location of injection ports shown by black stars in \textbf{d}. $D = 10^{-9}$ m$^2$/s. \textbf{a}-\textbf{c}: End of injection in lower port. A contour at $C_{\text{CO}_2} = 0.15\times C_{\text{CO}_2}^{\text{max}}\approx 0.21$ kg/m$^3$, the value used to compute areas with dissolved CO$_2$, is shown as a white solid line in \textbf{c}. \textbf{d}-\textbf{f}: End of injection in upper port. \textbf{g}-\textbf{i}: Time at which the first finger touches the rig bottom. \textbf{j}-\textbf{l}: Time at which the first finger touches the middle C layer. Colormaps used in the simulations by~\cite{thyng2016}.}
\label{fig:match}
\end{figure}

Fig.~\ref{fig:match} shows gas saturation ($S_\text{g}$) and CO$_2$ concentration ($C_{\text{CO}_2}$) maps at times at which quantities 1-4 described in sect.~\ref{sec:calibration} are evaluated. Snapshots are provided for model III only, since all three calibrated models were qualitatively very similar. It can be seen that CO$_2$ migration is successfully approximated by our numerical model. In detail, however, some differences are apparent: Firstly, sinking of CO2-rich water from the bottom injector and horizontal migration along the bottom of the rig is faster in the model. This is due to the higher permeability that our numerical model requires in order to match the gravity fingering advance (cf. Tab.~\ref{tab:matchedProperties}). Secondly, the experiment shows that denser, CO$_2$-rich water sinks with a rather compact front and closely spaced, wide fingers. Our model with constant $D = 10^{-9}$ m$^2$/s approximates all gravity-driven migration of the CO$_2$-rich water through thinner fingers, with the CO$_2$-saturated region receding with $S_\text{g}$. To better represent fingering widths, we also matched model III with $D = 3\times10^{-9}$ m$^2$/s, used in sect.~\ref{sec:benchmark}.

\subsection{Transferability: model predictions} \label{sec:predictions}
A key question after history matching a flow simulation model is whether the physical description has actually been improved, or whether parameters have been modified to match a set of specific observations only. By applying the history-matched models to a different injection protocol (experiment A2 in tank 1; refer to Tab.~\ref{tab:experimentSchedule}), and subsequently to a different geometry (experiment B1 in tank 2), this can be assessed to some extent. 

\subsubsection{Analog for a longer CO$_2$ injection in the same geologic setting} \label{sec:a2}
This case illustrates performance of our history matched models in a much longer injection in the same geology (experiment A2). Before simulating this case, we observed that the trapped gas column against the fault in the experiment was different than what could be achieved with our previous $p_\text{e}$ for models I-III (Tab.~\ref{tab:matchedProperties}). Because the capillary properties of the C sand in the fault were not directly involved in experiment A1, we increased $p_\text{e}$ in our calibrated models for that specific region ($p_\text{e}$ = 5 mbar against the lower F sand, and 3.5 mbar against the middle F sand). All other parameters were taken from the values calibrated to match experiment A1.

\begin{figure}[h]%
\centering
\includegraphics[width=1\textwidth]{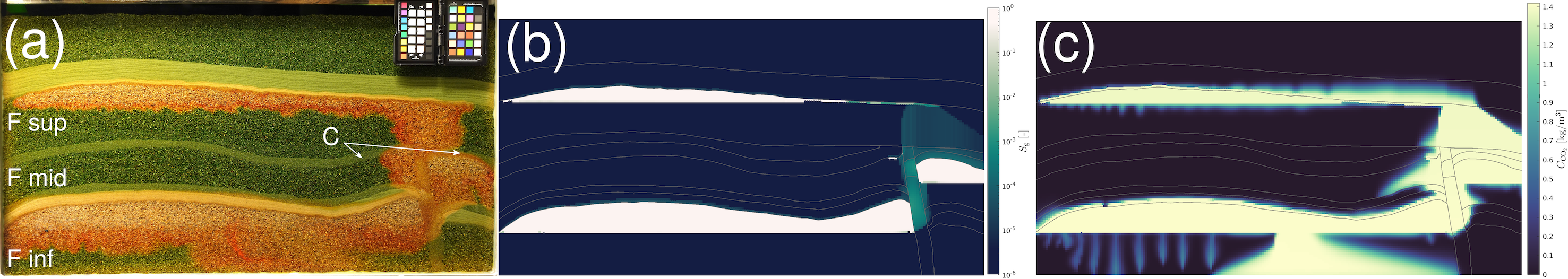}
\caption{Comparison between experiment A2 in tank 1 (\textbf{a}) and simulation results with model III (\textbf{b,c}) at the end of the injection phase ($t=4$h 48 min).}
\label{fig:AC07_match}
\end{figure}

Evaluation was performed at the end of injection, at $t = 4$h 48min, with a single run with models I-III. $I_\text{R}$ and $D$ were set to the same value in all three models: 1.7 ml/min and $10^{-9}$ m/s$^2$, respectively. The experimental result is shown in Fig.~\ref{fig:AC07_match}a, while the simulation with model III is depicted in Fig.~\ref{fig:AC07_match}b,c. We observe that the general distribution of CO$_2$ is close to the experimental truth. However, the experiment shows a compact sinking front of the CO$_2$-rich water without fingers; in our model, gravity fingering is apparent at this stage and fingers are close to the bottom of the rig. Additionally, CO$_2$-saturated brine touches the right boundary in the upper F reservoir, which does not occur in the experiment. This is due to capillary breach of the C sand above the middle F reservoir, as shown in Fig.~\ref{fig:AC07_match}b, and can be avoided by reducing the gas saturation value at which $p_\text{e}$ is defined, or by increasing $p_\text{e}$.

\begin{figure}[h]%
\centering
\includegraphics[width=0.5\textwidth]{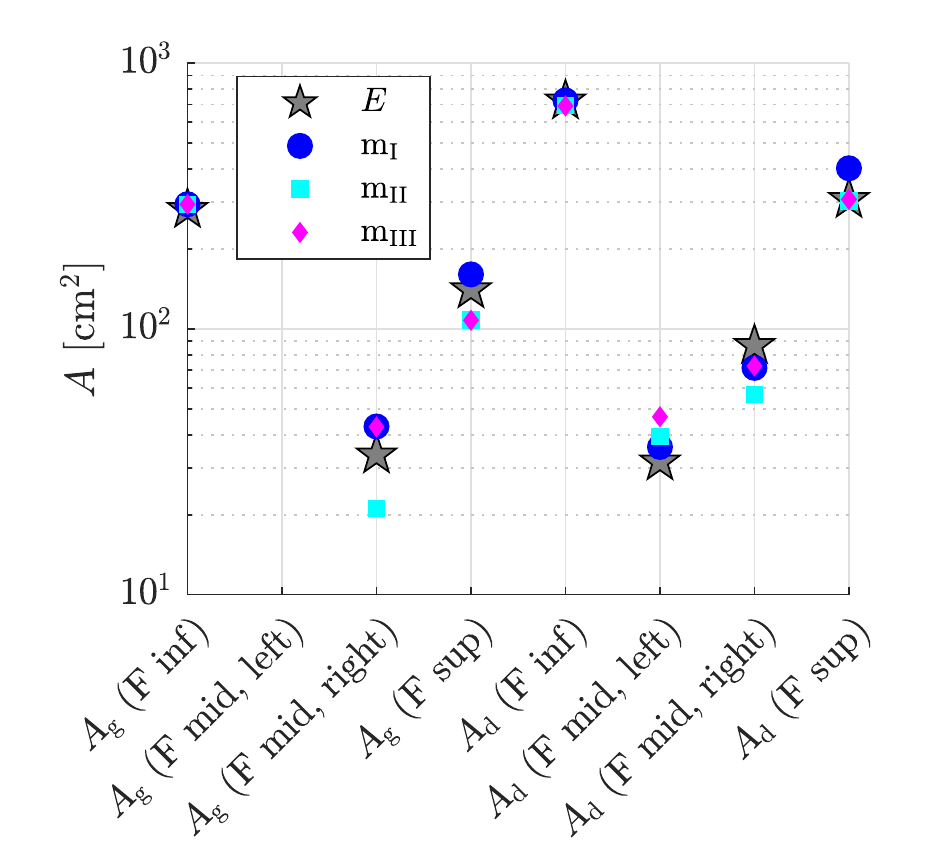}
\caption{Comparison of areas occupied by free gas ($A_\text{g}$) and water with dissolved CO$_2$ ($A_\text{d}$) for experiment A2 in tank 1. Experimental reference shown with a star ($E$). $A_\text{g}$ (F mid, left) not shown because values are very close to 0. Refer to Fig.~\ref{fig:experimentalvolume} or Fig.~\ref{fig:AC07_match}a for region location.}
\label{fig:AC07_areas}
\end{figure}

The comparison of areal quantities is provided in Fig.~\ref{fig:AC07_areas}, and shows good agreement between the experiment and simulation models. Model II (MAE = 16 cm$^2$) and III (MAE = 14.54 cm$^2$) perform similarly and slightly better than model I (MAE = 20.18 cm$^2$), but there are no marked differences. 

\subsubsection{Analog for a larger-scale CO$_2$ injection in a different geologic setting} \label{sec:benchmark}
Finally, we compare the performance of our calibrated models against experiment B1, conducted in a larger-scale, more complex geology (Fig.~\ref{fig:experimentalSetup}e)~\citep[this issue]{flemisch2023}. Similar to sect.~\ref{sec:a2}, our goal is to assess the predictive capability of our calibrated models---without changing their properties. However, given that sand D controls migration in the lower fault (see Fig.~\ref{fig:grid}e) and it was not present in our calibrated models, we allowed one change for models I and II, which did not have access to local $p_\text{c}$ measurements. This means that we ran an initial simulation of this experiment with model I and II, and then adjusted the $p_\text{c}$ curve of the D sand. The selected curve lies at $\approx \frac{1}{3}$ of the $p_\text{c}(S_\text{w})$ shown in Fig.~\ref{fig:krpc} and Fig.~\ref{fig:krpc2}, respectively. 

Next, we evaluated the performance of models I-III by comparing them to the experimental truth after a single run. Evaluation is performed over the total duration of the experiment (120 h), which is simulated with the same $I_\text{R}$ (10 ml/min) and $D$ ($10^{-9}$ m$^2$/s) in all three models. Additionally, a run with $D=3\times10^{-9}$ m$^2$/s was completed with model III to better approximate finger widths, as noted in sect.~\ref{sec:hm}. 

\begin{figure}[h!]%
\centering
\includegraphics[width=1\textwidth]{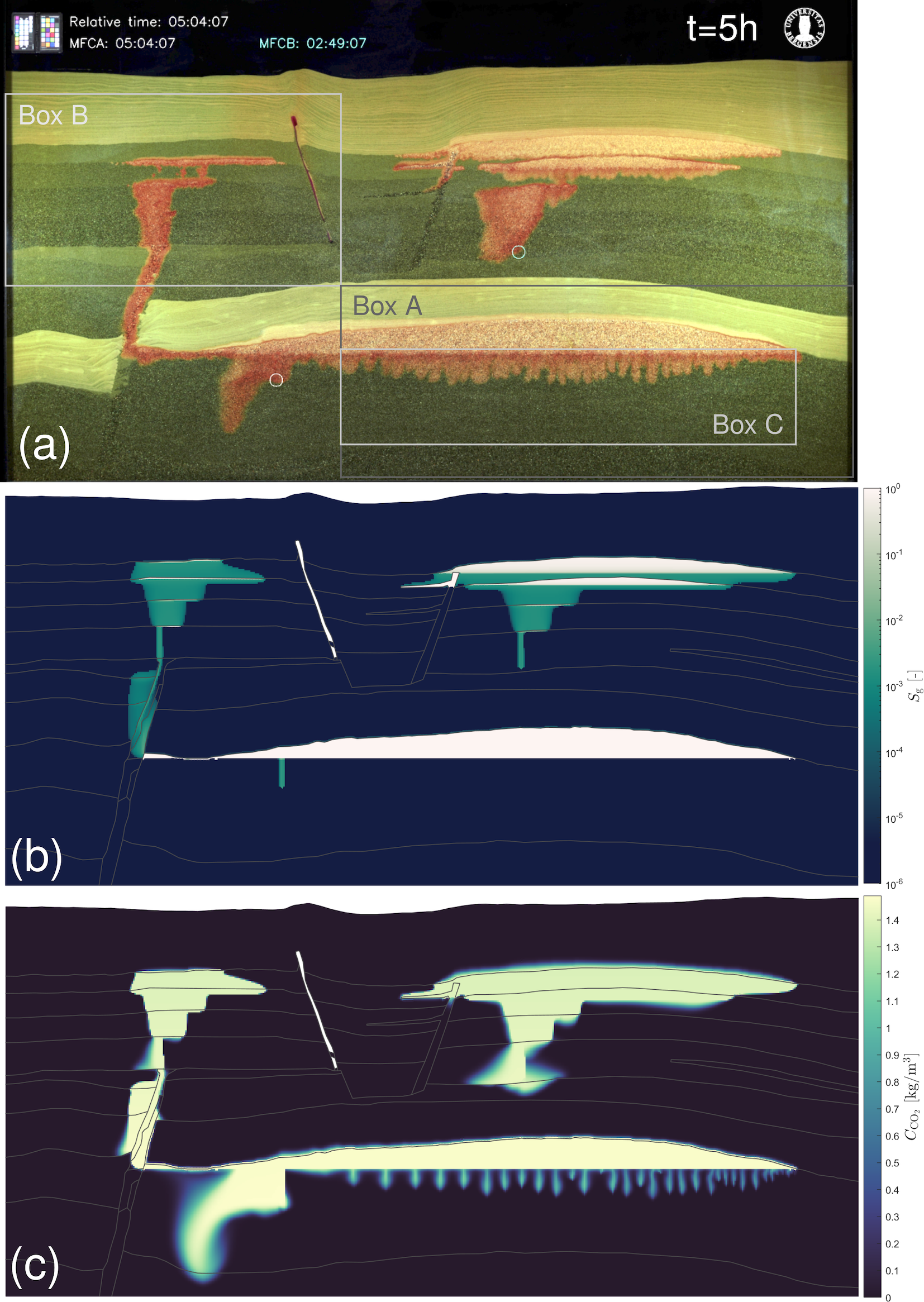}
\caption{Comparison between experiment B1 in tank 2 (\textbf{a}) and simulation model I (\textbf{b,c}) at the end of injection ($t = 5$h). Circles in \textbf{a} denote the location of injection ports.}
\label{fig:bm_match_5h}
\end{figure}

\begin{figure}[h!]%
\centering
\includegraphics[width=1\textwidth]{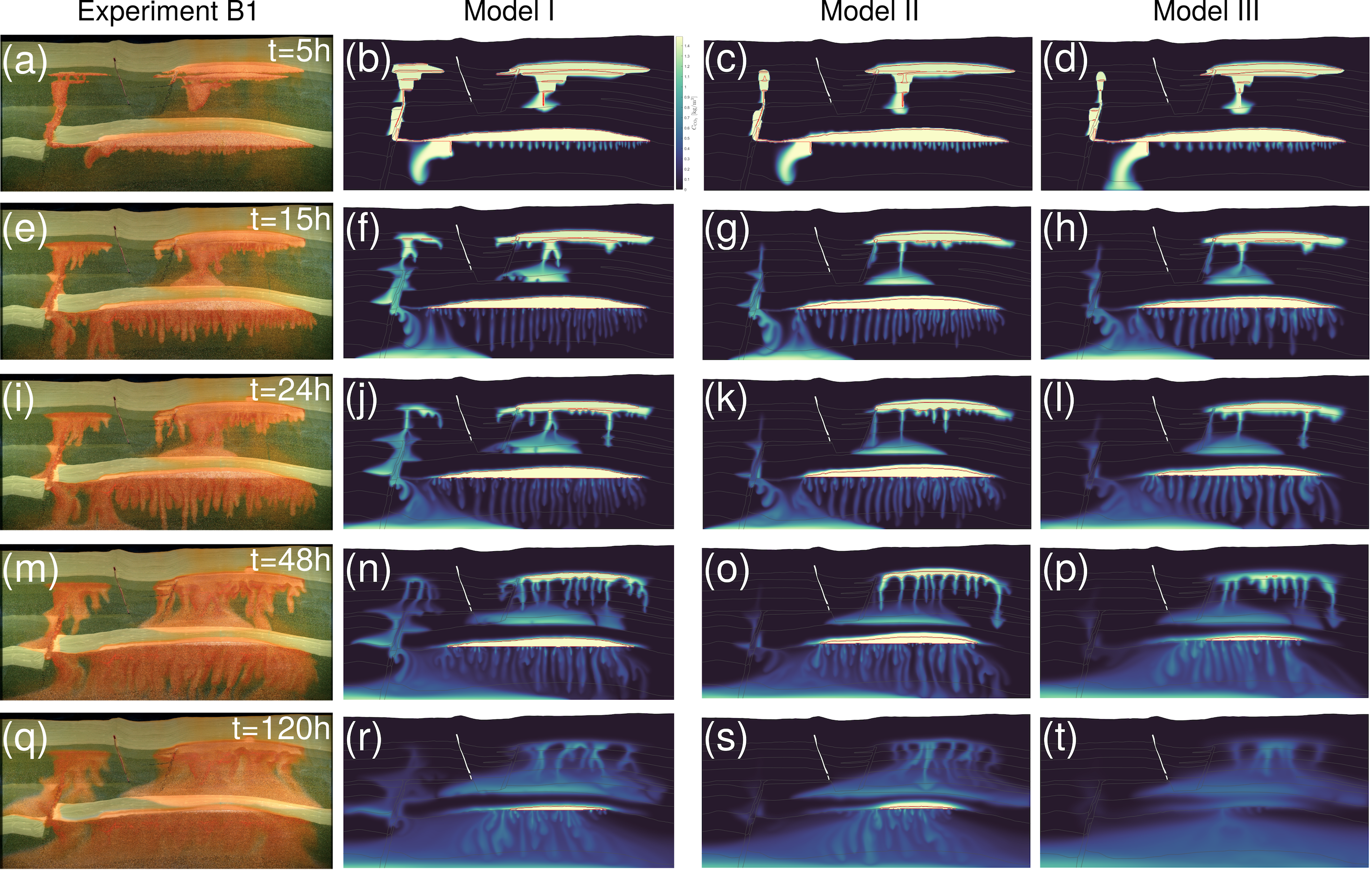}
\caption{Comparison between experiment B1 in tank 2 (leftmost column) and CO$_2$ concentration maps for simulation models I-III (middle-left, middle-right and rightmost, respectively). $D = 10^{-9}$ m$^2$/s (model I and II), $D = 3\times10^{-9}$ m$^2$/s (model III). The red contours in simulation plots indicate $S_\text{g} = 10^{-3}$. \textbf{a}-\textbf{d} end of injection. \textbf{e}-\textbf{h} $t=15$h. \textbf{i}-\textbf{l} $t=24$h. \textbf{m}-\textbf{p} $t=48$h. \textbf{q}-\textbf{t} $t=120$h.}
\label{fig:bm_match}
\end{figure}

Gas saturation and CO$_2$ concentration maps at the end of injection with model I are shown in Fig.~\ref{fig:bm_match_5h}a and Fig.~\ref{fig:bm_match_5h}b, respectively. The full visual comparison is provided in Fig.~\ref{fig:bm_match}. We make the following observations:
\begin{itemize}
    \item At the end of injection ($t = 5$ h), all three models predict some migration of CO$_2$ into box B. Model II (Fig.~\ref{fig:bm_match}c) and III (Fig.~\ref{fig:bm_match}d) underestimate the amount of CO$_2$, while model I (Fig.~\ref{fig:bm_match}b) overestimates the amount of CO$_2$ in the top C sand.
    \item Also at the end of injection, all models predict faster sinking of the CO$_2$-charged water tongue arising from the lower injector. This is due to the higher F sand permeability required to match finger advance (see sect.~\ref{sec:hm}), particularly in model III with $D = 3\times10^{-9}$ m$^2$/s.
    \item The speed at which CO2-rich fingers sink is slightly faster in our models, compared to the experiment. As expected, model III, with a higher diffusion coefficient, displays thicker fingers, with closer widths to the experiment. Similar to our previous observations, the numerical models cannot approximate the compact, CO$_2$-rich water front closely trailing the fingers.
    \item Dissolution of CO$_2$ is underestimated by models I and II, while it is closer, but overestimated, by model III.
\end{itemize}

\begin{figure}[h]%
\centering
\includegraphics[width=1\textwidth]{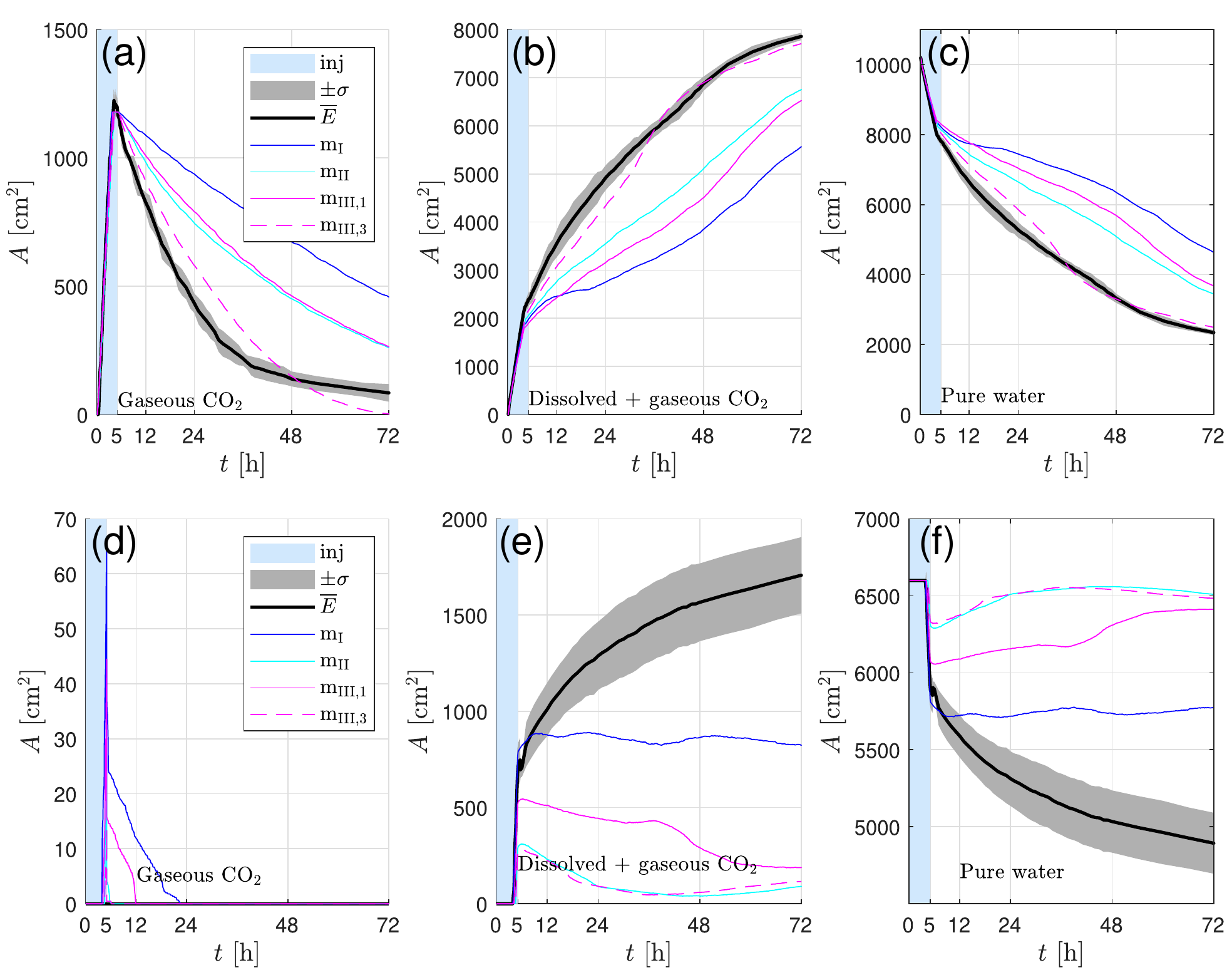}
\caption{Comparison of areas occupied by each phase during the first 72h of case B1. Experimental mean ($\overline{E}$) and standard deviation ($\sigma$) obtained from four experimental runs with identical protocol, while the results for models I-III are for a single run with each matched model. For m$_\text{III}$, two cases are shown: $D = 10^{-9}$ m$^2$/s (m$_{\text{III},1}$) and $D = 3\times10^{-9}$ m$^2$/s (m$_{\text{III},3}$). Top row shows areas in box A, and bottom row shows areas in box B. \textbf{a,d} gaseous CO$_2$. \textbf{b,e} dissolved CO$_2$ (includes area with gaseous CO$_2$). \textbf{c,f} pure water.}
\label{fig:bm_areas}
\end{figure}

Consistent with our approach described in sect.~\ref{sec:calibration}, quantitative analysis is provided by means of areal quantities over time in Fig.~\ref{fig:bm_areas}. Experimental values were obtained via segmentation of timelapse images, and the data was reported on a 1$\times$1 cm grid where 0 is pure water, 1 is water with dissolved CO$_2$, and 2 is gaseous CO$_2$. The segmentation procedure is explained in~\cite{nordbotten2023}, this issue. We then obtained the areas of each phase within box A and B to generate Fig.~\ref{fig:bm_areas} (refer to Fig.~\ref{fig:bm_match_5h}a for box location). 

In box A, which contains the main F reservoir and ESF seal, we observe very accurate areas during injection. Afterwards, model III with $D = 3\times10^{-9}$ m$^2$/s continues to follow the experiment closely, whereas the others overestimate gaseous CO$_2$. Note that the PVT properties of our fluids are the same in all models; differences arise due to (1) higher sand F $S_\text{wc}$ in model III, and higher sand F $k$ in model II and especially model III ($D = 3\times10^{-9}$ m$^2$/s), compared to model I, which allow greater convective mixing~\citep{ennis2005}(Tab.~\ref{tab:matchedProperties}); and (2) lower $p_\text{e}$ and higher $k$ of sand ESF in model II (Tab.~\ref{tab:matchedProperties}), which allows some CO$_2$ migration into the seal (Fig.~\ref{fig:bm_match}). In box B (Fig.~\ref{fig:bm_areas}d-f), model I and model III with $D = 10^{-9}$ m$^2$/s are able to approximately track the experimental truth during injection. However, our models without dispersion cannot capture the areal increase of CO$_2$-rich water that occurs afterwards (cf. Fig.~\ref{fig:bm_match}).

Further evaluation of simulation model performance, including comparison of mass quantities and error measures, is provided in Appendix~\ref{sec:supplAnalysisbm}. From this analysis (sect.~\ref{sec:predictions} and Appendix~\ref{sec:supplAnalysisbm}), we find that all three calibrated models provide a reasonable prediction of the experiment. Qualitatively, they approximate well CO$_2$ migration, seal capacity, and onset of convective mixing. Quantitatively, our models are able to accurately estimate specific quantities during the injection phase, yet they accumulate higher errors later on (Fig.~\ref{fig:bm_areas} and Appendix~\ref{sec:supplAnalysisbm}). Overall, model III performs marginally better than model I and II (Appendix~\ref{sec:supplAnalysisbm}), but, similar to sect.~\ref{sec:a2}, the differences are not marked or consistent. The system is very sensitive to variations in petrophysical properties such as capillary pressure; therefore, in regions dominated by structures with uncertain properties (like box B in experiment B1), our estimates were less accurate. 

In summary, calibrated models are transferable to a different operational setting or geologic structure, as long as sediments and trap systems remain similar (experiment A2 and box A in experiment B1). Where reservoir connectivity is provided by heterogeneous structures with uncertain properties, accurate deterministic estimates of CO$_2$ migration are unlikely; models calibrated elsewhere (experiment A1) did not perform well in our test (box B in experiment B1). Given unlimited computational time, the predictive ability of numerical models calibrated with published data appears similar to those having access to local measurements. The main value of local data lies in reducing the time required for history matching.

\section{Discussion}\label{sec:discussion}
Our experiments and numerical simulations of CO$_2$ injection and migration in unconsolidated sands at atmospheric $p,\, T$ conditions capture the CO$_2$-water system dynamics at short to intermediate timescales: buoyancy-driven flow and structural trapping~\citep{bachu1994,bryant2008,hesse2010,szulczewski2013}, residual trapping~\citep{juanes2006,burnside2014} and convective mixing and dissolution trapping~\citep{weir1996,ennis2005,riaz2006,neufeld2010,hidalgo2012,macminn2013,szulczewski2013}. Due to the very large sand permeability ($10^2-10^4$ D), the latter dominates CO$_2$ trapping. 

The strong buoyancy and high permeability lead to persistent appearance and disappearance of fluid phases, as the gas migrates upward and dissolves in the water; this renders the nonlinear solver unstable, and, coupled with the nonlinearities introduced by relative permeability hystreresis, makes this problem difficult to solve numerically~\citep[e.g.,][]{lie19}. As hinted in sect.~\ref{sec:setup}, this was addressed by optimizing linear solver time, reducing the timestep length, increasing the number of timestep cuts and relaxing MRST's maximum normalized residual where required.

In a 2D isotropic medium and assuming uniform flow, the hydrodynamic dispersion coefficient ($\underline{\underline{D}}_\text{h}$) can be modeled as $\underline{\underline{D}}_\text{h} = \bigl[ \begin{smallmatrix}\alpha_\text{L}\overline{u} & 0\\ 0 & \alpha_\text{T}\overline{u}\end{smallmatrix} \bigr]$, where $\alpha_\text{L}$ and $\alpha_\text{T}$ are the longitudinal and transverse dispersivity, respectively, and $\overline{u}$ is the average Darcy velocity~\citep{bear1972}. Assuming dispersivities $\geq 10^{-2}-10^{-3}$ m~\citep{garabedian1991,gelhar1992,schulze2005} and $\overline{u}\approx 3\times10^{-6}$ m/s (from our simulations), we get $\underline{\underline{D}}_\text{h} \in [3\times10^{-9}, \, 3\times10^{-8}]$ m$^2$/s or larger; this means that $\underline{\underline{D}}_\text{h} \geq D$ for the timescales considered~\citep{riaz2004,rezk2022}. Previous work suggested that hydrodynamic dispersion in homogeneous sediments can be accounted for by increasing $D$~\citep{riaz2004,riaz2006}, as done here. However, our analysis shows that the spreading of CO$_2$-rich water during convective mixing can be loosely, but not accurately, represented by molecular diffusion. Given (1) the dominance of convective mixing on solubility trapping~\citep{ennis2005,neufeld2010,macminn2013}; (2) heterogeneity of many natural reservoirs, which increases the importance of dispersion~\citep{riaz2006,bear2018}; and (3) the acceleration of CO$_2$ dissolution due to dispersion, as observed here and by others~\citep[e.g.,][]{hidalgo2009}, it is important to quantify the balance between diffusion and dispersion to estimate CO$_2$ trapping.

The models developed in this paper retained some error at the end of the calibration phase, which is a known problem of manual history matching~\citep{oliver2011}. This approach was chosen given the low geometric uncertainty; the availability of both local petrophysical measurements and the experimental truth; and the long computational times and number of uncertain variables. Consistent with previous results~\citep[e.g.,][]{fisher2007}, results show that model II and III, which had access to local petrophysical measurements, achieved faster match to the experimental truth than model I (sect.~\ref{sec:hm}). However, all models were similarly accurate when making predictions (sect.~\ref{sec:predictions}). Subsurface formations are significantly more heterogeneous than the silica sands employed here; together with time constraints, this may explain why existing literature emphasizes the importance of accurate geologic representation and reservoir model parametrization, including local data, to achieve successful history matching, and, especially, forecasting~\citep[e.g.,][]{gosselin2003,fisher2007,myers2007,kam2015,avansi2016}. 

Because of the availability of a ground truth to assess model predictions, we did not quantify uncertainty in the history matched models here. In general, however, this is necessary to manage reservoir operations~\citep[e.g.,][forthcoming, and references therein]{aanonsen2009,oliver2011,jagalur2018,jin2019,liu2020,santoso2021,landa2023}. It is also important to note that history-matched models may have grid-size dependencies (see Appendix~\ref{sec:h}), which may require that the grid used to make predictions, if different or encompassing additional regions, maintain a similar resolution. Finally, multiphase flow in poorly-lithified sediments is non-unique~\citep[this issue]{haugen2023}, which also contributes to uncertainty. Due to the sensitivity of the CO$_2$-brine system to relatively small reservoir parameter changes, as observed in sect.~\ref{sec:benchmark}, it seems prudent to adopt a probabilistic perspective when estimating subsurface CO$_2$ migration~\citep[see][this issue]{flemisch2023}.

\section{Conclusions}\label{sec:conclusions}
We performed experiments (sect.~\ref{sec:experiments}) and numerical simulations (sect.~\ref{sec:numerical}) of CO$_2$ migration in poorly-lithified, siliciclastic sediments at the meter scale. Three simulation models, with access to different levels of local data, were manually history-matched to the experiments (sect.~\ref{sec:run1},~\ref{sec:hm}), and then used to make predictions (sect.~\ref{sec:predictions}). The main findings are:
\begin{enumerate} 
    \item The time required to history match model III (access to both single-phase and multiphase measurements) is lower than model II (access to local single-phase measurements), which is lower than model I (no access to local petrophysical measurements).
    \item All simulation models achieve a satisfactory qualitative match throughout the experiments. Quantitatively, overall prediction accuracy of models I-III is similar: our models are close to the experimental truth during CO$_2$ injection, and accumulate larger errors afterwards, especially in regions where heterogeneous structures control CO$_2$ migration.
    \item The addition of a constant molecular diffusion coefficient allows matching convective finger widths to experimental observations. However, simulations without dispersion cannot approximate the compact, CO$_2$-rich sinking front closely trailing convective fingers in our experiments. 
\end{enumerate}
Simulation models were not always accurate. Given the degree of control in our study, it seems prudent to quantify uncertainty when assessing subsurface CO$_2$ migration in the field using numerical models. Obtained results suggest that confidence can be increased by obtaining local data (time is finite), quantifying petrophysical parameter uncertainty, testing sensitivity to petrophysical parameters in different model regions, including post-injection data when history matching, and incorporating multiple scenarios of CO$_2$ migration, particularly where heterogeneous structures are at play.

\backmatter


\bmhead{Acknowledgments}
LS gratefully acknowledges laboratory support provided by UiB Engineer Emil Bang Larsen and image processing support provided by UiB PhD student Benyamine Benali. LS and RJ are grateful to Olav Møyner and the MRST team at SINTEF for their guidance to implement new functionality and continuous support with the MATLAB Reservoir Simulation Toolbox.  A special thanks goes to Robert Gawthorpe, Atle Rotevatn and Casey Nixon for helpful comments on the geology of North Sea reservoirs, which were key to build the geometry in tank 2. The authors also acknowledge support by the following organizations (refer to `Funding' below for details): ExxonMobil, ``la Caixa" Foundation, Research Council of Norway (RCN), Akademia.

\section*{Statements and Declarations} \label{sec:declarations}

\begin{itemize}
\item \textbf{Funding} This work was supported by ExxonMobil through the project ``Modeling and Mitigation of Induced Seismicity and Fault Leakage during CO$_2$ storage". LS gratefully acknowledges the support of a fellowship from ``la Caixa" Foundation (ID 100010434). The fellowship code is LCF/BQ/EU21/11890139. MH is funded by the Research Council of Norway (RCN) project no. 280341. KE, MH, JMN and MF are partly funded by the Centre for Sustainable Subsurface Resources, RCN project no. 331841. The work of JMN and MF is funded in part through the Akademia project ``FracFlow".

\item \textbf{Competing interests} The authors declare that they have no conflict of interest.
\item \textbf{Availability of data and materials} The dataset for experiment B1 can be obtained from \href{github.com/fluidflower}{https://github.com/fluidflower}. The remaining experimental and simulation data are available from the corresponding authors on reasonable request.
\item \textbf{Authors' contributions} JMN, MF and RJ designed the study and acquired the funding. JMN, MF and KE conceptualized, designed and built the FluidFlower rigs. MH, KE and MF conducted the ex-situ sand property measurements. LS designed the experiments in tank 1. JMN, MF, MH and KE designed the experiments in tank 2. MH, KE, LS and MF conducted the experiments. LS developed the simulation models and conducted the simulations. LS, JMN and RJ performed the simulation analysis. LS wrote the paper, with inputs from all authors.
\end{itemize}

\begin{appendices}

\section{Additional analysis of simulation model performance against experiment B1} \label{sec:supplAnalysisbm}
\begin{figure}[h]%
\centering
\includegraphics[width=1\textwidth]{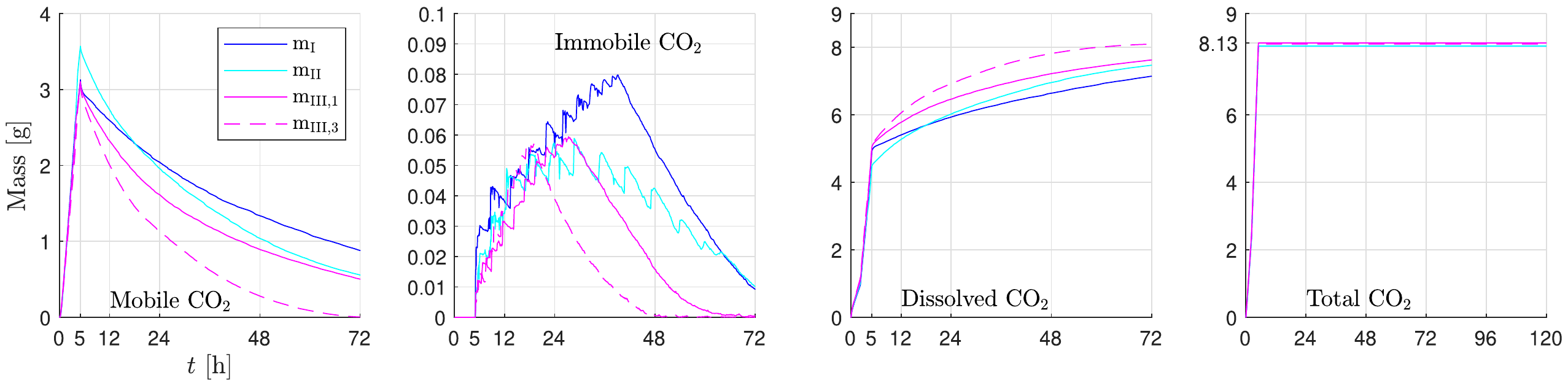}
\caption{Total mass of CO$_2$ for our simulations of experiment B1 presented in sect.~\ref{sec:benchmark}. Results are provided for models I to III. For m$_\text{III}$, two cases are shown: $D = 10^{-9}$ m$^2$/s (m$_{\text{III},1}$) and $D = 3\times10^{-9}$ m$^2$/s (m$_{\text{III},3}$).}
\label{fig:bm_mass}
\end{figure}

\begin{figure}[h]%
\centering
\includegraphics[width=1\textwidth]{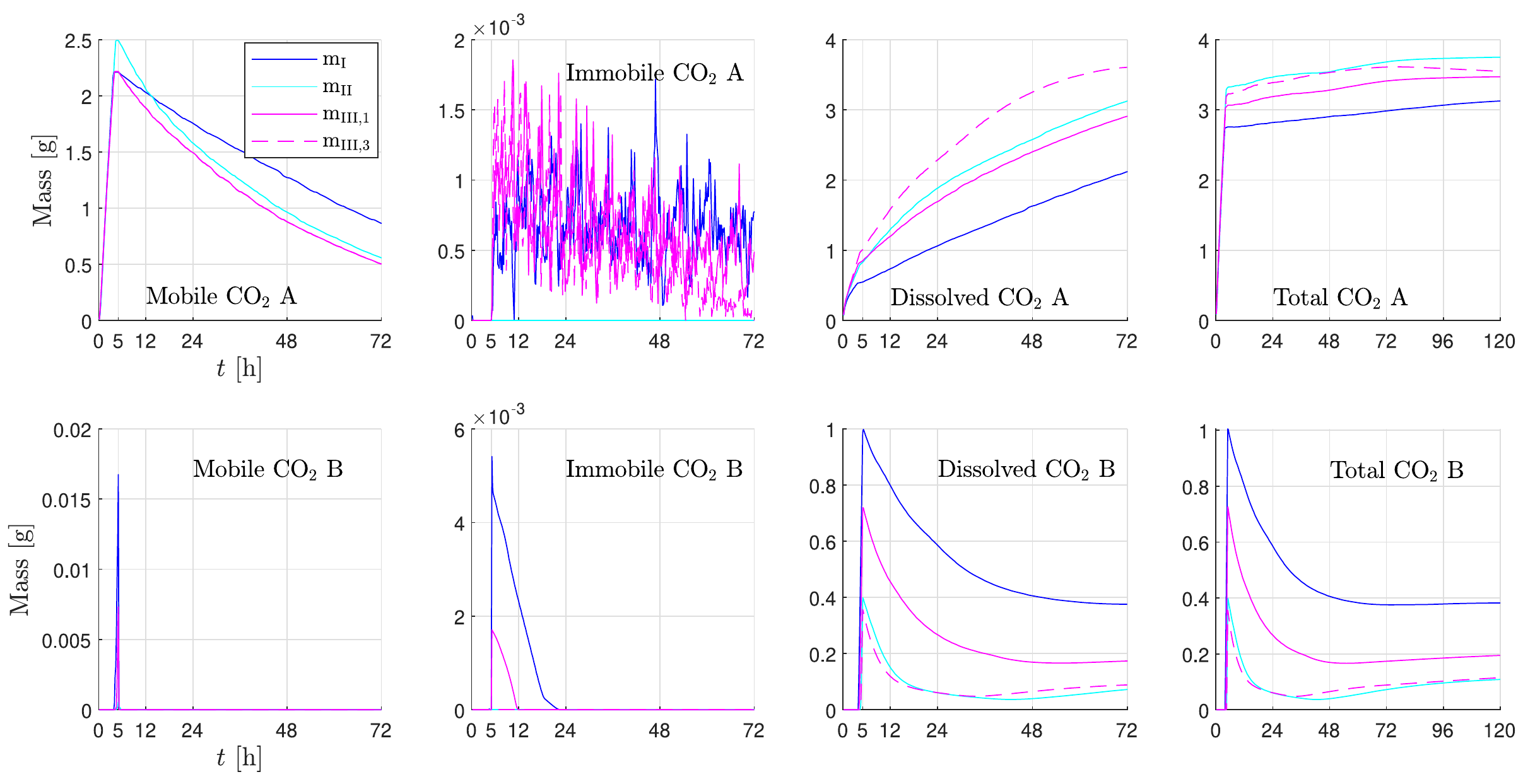}
\caption{Mass of CO$_2$ in boxes A and B defined in Fig.~\ref{fig:experimentalSetup}e, for our simulations of experiment B1 presented in sect.~\ref{sec:benchmark}. Results are provided for models I to III. For m$_\text{III}$, two cases are shown: $D = 10^{-9}$ m$^2$/s (m$_{\text{III},1}$) and $D = 3\times10^{-9}$ m$^2$/s (m$_{\text{III},3}$).}
\label{fig:bm_mass_box}
\end{figure}

First, we provide the total mass of CO$_2$ in the computational domain in Fig.~\ref{fig:bm_mass}, and the mass in boxes A and B in Fig.~\ref{fig:bm_mass_box}.

Next, in Tab.~\ref{tab:sparse}, the following measures are compared with quantities estimated from the experiment via segmentation of timelapse images~\citep{nordbotten2023}. These measures correspond to the sparse data requested to participants of the FluidFlower international benchmark study~\citep{flemisch2023}: 
\begin{enumerate}
  \item time of maximum mobile free phase in box A
  \item mass of mobile CO$_{2(\text{g})}$, immobile CO$_{2(\text{g})}$, dissolved CO$_2$, and CO$_2$ in the seal (in any phase), in box A, 72 h after injection start (2a-d)
  \item the same quantities as 2. for box B (3a-d)
  \item time at which $M$ (defined below) exceeds 110\% of the width of box C
  \item total mass of CO$_2$ in the ESF seal, in box A, at $t = 120$ h
\end{enumerate}

Convective mixing in box C (see Fig.~\ref{fig:experimentalSetup}e) is reported as the integral of the magnitude of the gradient in relative concentration of dissolved CO$_2$~\citep{flemisch2023}:
\begin{equation}
    M(t) = \int_\text{C} \bigg\lvert \nabla\bigg(\frac{\chi_{\text{CO}_2}^\text{w}}{\chi_{\text{CO}_2}^\text{w,max}}\bigg)\bigg\rvert d\mathbf{x}
    \label{eq:M}
\end{equation}
where $\chi_{\text{CO}_2}^\text{w}$ is the mass fraction of CO$_2$ in water, and the dissolution limit is $\chi_{\text{CO}_2}^\text{w,max}$. Note that quantity 4, based on $M$, cannot be provided with full accuracy based on experimental data, so an uncertain lower and upper bound is provided instead. Therefore, we assigned an error ($\varepsilon$) = 0 in Tab.~\ref{tab:sparse}. 

Relative error is evaluated with respect to the experimental mean ($\overline{E}$) as $\varepsilon_i (\%) = 100\times\frac{\lvert \overline{E_i} - \text{m}_{J,i} \rvert}{\overline{E_i}}$, where $i$ is a given measure and $J$ refers to any of the models I-III. In Tab.~\ref{tab:sparse}, it can be seen that all models accumulate some error in most of the quantities reported. The maximum errors are $\approx 140\%$ for models I-II and $< 100\%$ for model III. Model I performs better in the uncertain region (box B; see sect.~\ref{sec:benchmark} as well), while models II and III are more accurate in box A, the region where the calibration performed with experiment A1 is more meaningful. Overall, model III does marginally better.

\begin{sidewaystable} 
\centering
\caption{Sparse data comparison between experiment B1 in tank 2 and simulation results with models I-III. Experimental mean and standard deviation were obtained from six experimental runs with identical protocol, while the results for models I-III are for a single run with each matched model. For m$_\text{III}$, two cases are shown: $D = 10^{-9}$ m$^2$/s (m$_{\text{III},1}$) and $D = 3\times10^{-9}$ m$^2$/s (m$_{\text{III},3}$). Experimental quantity 4 is reported using a lower and upper bound due to high uncertainty, so errors are not computed. See main text for measure description.}
\begin{tabular}{c c c c c c c c c c c}
\toprule
Measure & $\overline{E}$ & $\sigma (E)$ & m$_\text{I}$ & $\varepsilon_\text{I}$ [\%] & m$_\text{II}$ & $\varepsilon_\text{II}$ [\%] & m$_{\text{III},1}$ & $\varepsilon_{\text{III},1}$ [\%]  & m$_{\text{III},3}$ & $\varepsilon_{\text{III},3}$ [\%] \\ \midrule
  1 [s] & 14880 & 720 & 18000 & 21 & 17340 & 16.5 & 17880 & 20.2 & 17160 & 15.3\\
  2a [g] & 0.36 & 0.13 & 0.86 & 140.23 & 0.56 & 54.7 & 0.5 & 39.5 & 0.004 & 98.9\\
  2b [g] & 0 & 0 & 0 & 0 & 0 & 0 & 0 & 0 & 0 & 0\\
  2c [g] & 3.5 & 0.08 & 2.12 & 39.5 & 3.12 & 10.7 & 2.9 & 16.9  & 3.61 & 3.0\\
  2d [g] & - & - & 0.43 & n/a & 0.96 & n/a & 0.73 & n/a & 0.69 & n/a\\
  3a [g] & 0 & 0 & 0 & 0 & 0 & 0 & 0 & 0 & 0 & 0\\
  3b [g] & 0 & 0 & 0 & 0 & 0 & 0 & 0 & 0 & 0 & 0\\
  3c [g] & 0.55	& 0.32 & 0.38 &	31.7 & 0.072 & 86.9 & 0.17 & 68.5 & 0.09 & 83.9\\
  3d [g] & n/a & n/a & 0.0015 & n/a & 0.0053 & n/a & 0.0021 & n/a & 0.0039 & n/a\\
  4 [s] & [12180, 17990]	& [438, 2261] & 15000 & 0 & 15600 & 0 &	19200 & 0 & 15600 & 0\\
  5 [g] & 0.38 & 0.047 & 0.52 & 36.2 & 0.92 & 142.2 & 0.72 & 89.1 & 0.62 & 62.7\\ \bottomrule
  $\overline{\varepsilon}$ [\%] & n/a & n/a & n/a & 29.8 & n/a & 34.6 & n/a & 26.0 & n/a & 29.3\\ \bottomrule
\end{tabular}
\label{tab:sparse}
\end{sidewaystable} 

\section{Comparison of simulation results with multiple grid resolutions} \label{sec:h}
\begin{figure}%
  \centering
  \includegraphics[width=1\textwidth]{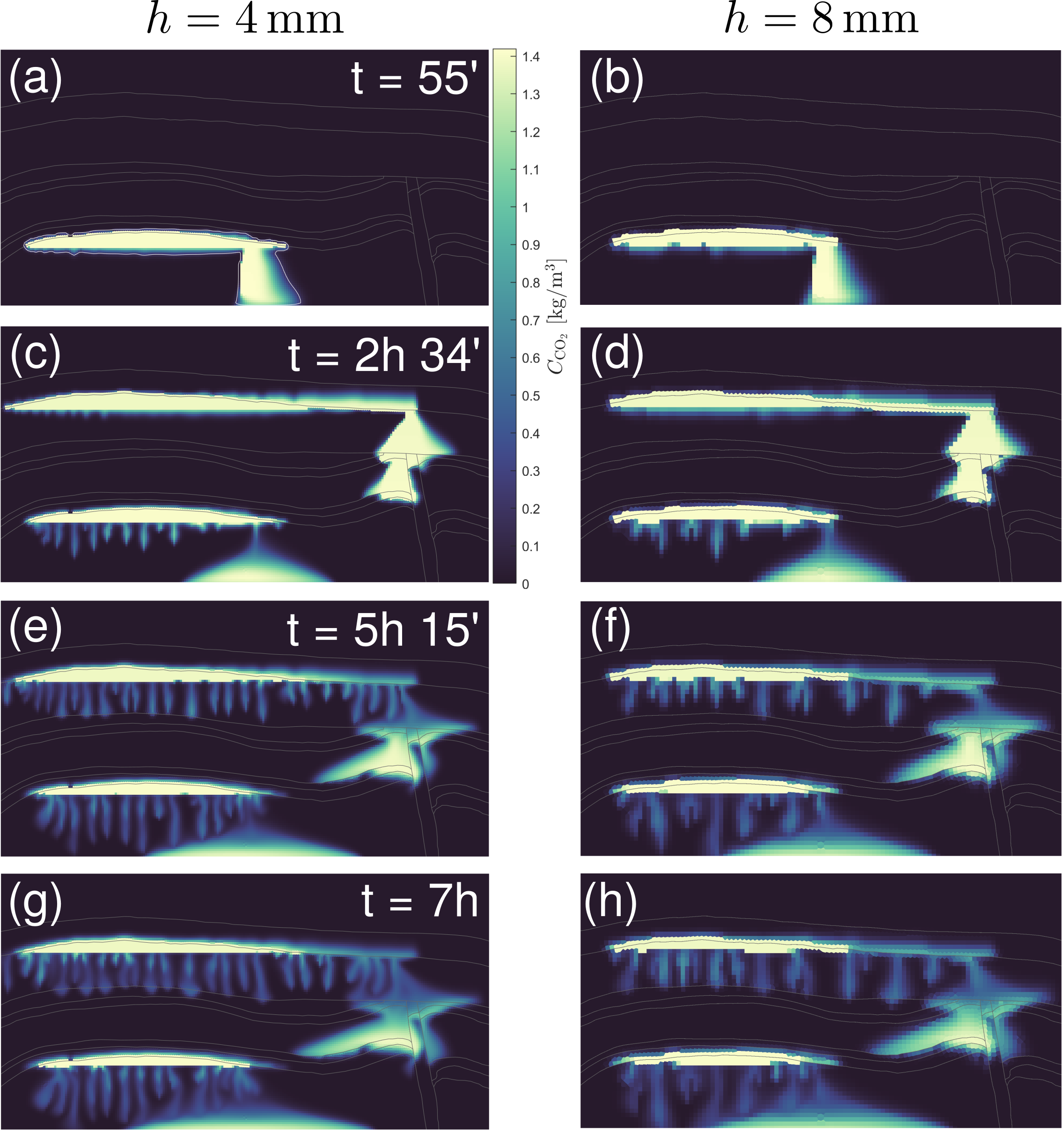}
  \caption{Concentration maps from our simulations of experiment A1 with model III. Results with two grids are shown: $h = 4$ mm (\textbf{a}, \textbf{c}, \textbf{e}, \textbf{g}) and $h = 8$ mm (\textbf{b}, \textbf{d}, \textbf{f}, \textbf{h}).}
  \label{fig:ac02_h}
\end{figure}

\begin{figure}%
  \centering
  \includegraphics[width=1\textwidth]{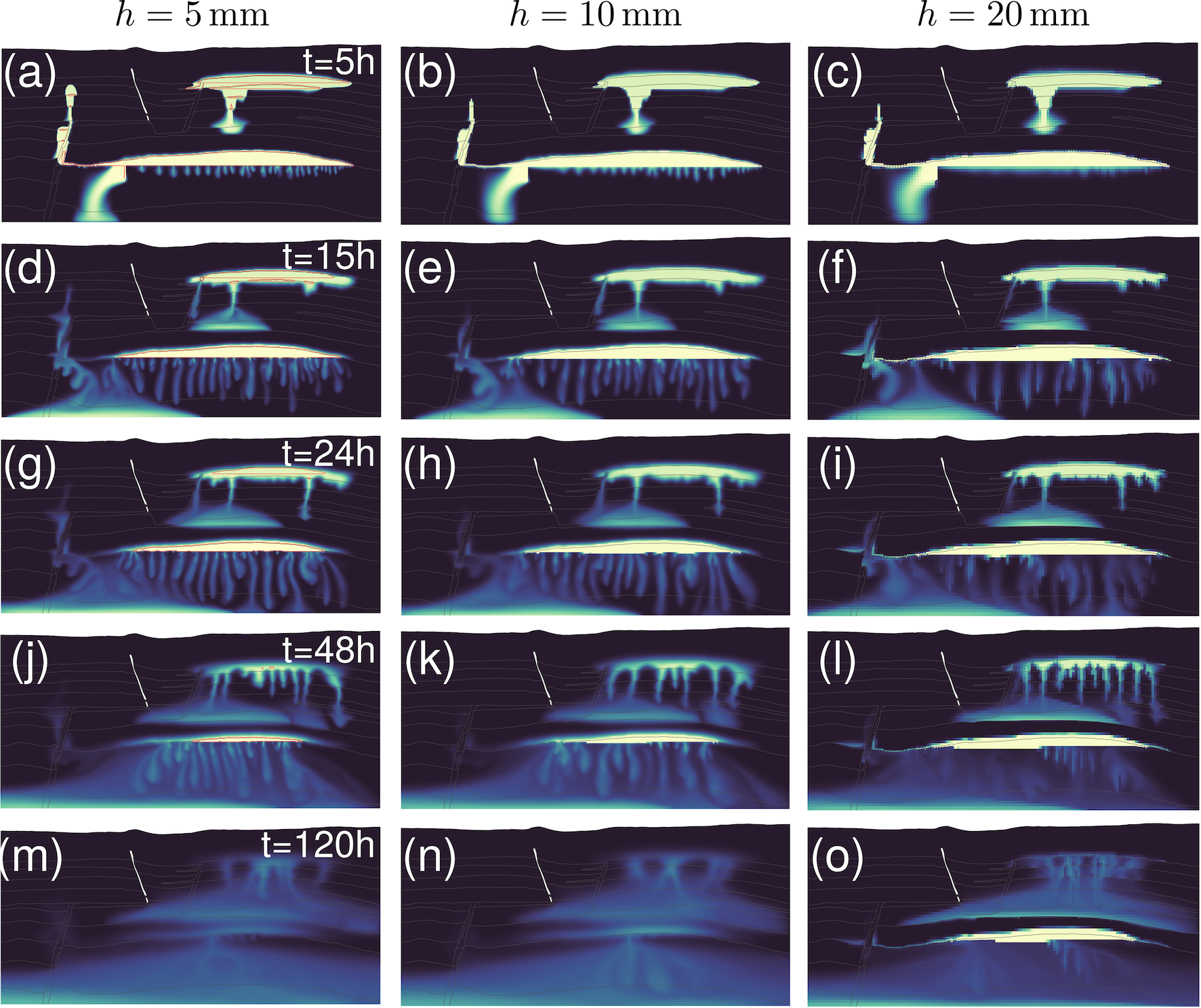}
  \caption{Concentration maps from simulations of experiment B1 with model III. Results with three grids are shown: $h = 5$ mm (\textbf{a}, \textbf{d}, \textbf{g}, \textbf{j}, \textbf{m}), $h = 10$ mm (\textbf{b}, \textbf{e}, \textbf{h}, \textbf{k}, \textbf{n}) and $h = 20$ mm (\textbf{c}, \textbf{f}, \textbf{i}, \textbf{l}, \textbf{o}).}
  \label{fig:bm_h}
\end{figure}

This section provides two comparisons of concentration maps obtained with model III after the calibration presented in sect.~\ref{sec:hm}:
\begin{enumerate}
  \item For experiment A1, we compare two grid sizes: $h = 4$ mm, as shown in the paper, and a coarser grid with $h = 8$ mm (Fig.~\ref{fig:ac02_h}).
  \item For experiment B1, we compare three grid sizes: $h = 5$ mm, as shown in the paper, and two coarser grids with $h = 10$ mm and $h = 20$ mm, respectively (Fig.~\ref{fig:bm_h}). 
\end{enumerate}
It can be seen that, for the calibrated parameter set (Tab.~\ref{tab:matchedProperties}), the coarser models maintain a general agreement with the finer ones (and the experimental solution). However, some differences are clear even in this qualitative comparison, including (1) smaller extent of the CO$_2$ plume, (2) lower dissolution, (3) lower number of fingers and finger widths, and (4) different CO$_2$-rich finger sinking speed. Therefore, the calibration process is somewhat cell-size dependent, which has implications for applying history matched models from e.g., pilot tests to field-scale CO$_2$ storage projects.

\end{appendices}

\clearpage
\bibliography{sn-article}%



\end{document}